\documentclass [] {aa}
\usepackage{epsfig}

\begin{document}

\thesaurus{05(10.15.2 Pleiades,08.18.1,08.05.3,08.01.02,03.20.8)}
\titlerunning{The rotational velocity of low-mass stars in the Pleiades}
\title{The rotational velocity of low-mass stars in the Pleiades
  cluster\thanks{Based on observations collected at the Observatoire de
    Haute-Provence with ELODIE at the 193cm telescope and with CORAVEL at
    the 1m-swiss telescope}} \author{D. 
Queloz\inst{1}\fnmsep\thanks{\emph{Present address:} Jet Propulsory Laboratory, 
Mail-Stop: 306-473, 4800 Oak Grove Drive, Pasadena, CA 91109, USA} \and S.
  Allain\inst{2} \and J.-C. Mermilliod\inst{3}\and J. Bouvier \inst{2}\and
  M. Mayor \inst{1}}

\institute{Observatoire de Gen\`eve, 51 ch. des Maillettes, CH--1290 Sauverny, 
Switzerland\and
           Laboratoire d"Astrophysique, Observatoire de Grenoble, Universit\'e 
Joseph Fourier, B.P. 53, 38041 Grenoble Cedex 9, France\and
           Institut d'Astronomie de l'Universit\'e de Lausanne, CH--1290 
Chavannes-des-bois, Switzerland}

\offprints{Didier.Queloz@obs.unige.ch}
\date{Received  January 98/ Accepted }
\maketitle

\begin{abstract}
We present new {$v\sin i$}~ measurements for 235 low-mass stars in the
Pleiades. The differential rotational broadening has been resolved for
all the stars in our sample. These results, combined with previously
published measurements, provide a complete and unbiased rotation data set
for stars in the mass range from 0.6 to 1.2{$M_{\odot}$}. Applying a numerical
inversion technique on the {$v\sin i$}~ distributions, we derive the
distributions of {\it equatorial\/} velocities for low-mass Pleiades
members. We find that half of the Pleiades dwarfs with a mass between 0.6
to 1\,{$M_{\odot}$}~ have rotation rates lower than 10{\,km\,s$^{-1}$}. \\
Comparison of the rotational distributions of low-mass members between IC
2602/2391 ($\approx 35$\,Myr) and the Pleiades ($\approx 100$\,Myr) suggests 
that G dwarfs behave like solid-bodies and follow Skumanich's law during this
time span. However, comparison between Pleiades and  older clusters
--M34 ($\approx 200$\,Myr) and Hyades ($\approx 600$\,Myr)--  indicates that 
the braking of slow rotators on the early main sequence is
weaker than predicted by an asymptotical Skumanich's law.  
This strongly supports the view that
angular momentum tapped in the radiative core of slow rotators on the
zero age main sequence (ZAMS) resurfaces into the convective envelope
between Pleiades and Hyades
age. For the G-dwarfs, we derive a characteristic coupling time scale
between the core and the envelope of about 100--200\,Myr, which accounts for
the observed evolution of surface rotation from the ZAMS to the Hyades.\\
The relationship between rotation and coronal activity in the Pleiades is
in agreement with previous observations in other clusters and field
stars.  We show that the Rossby diagram provides an excellent description
of the X--ray activity for all stars in the mass domain studied. The
Pleiades data for slow and moderate rotators fills the gap between the
X-ray--rotation correlation found for slow rotators and the X-ray
``saturation plateau'' observed for young fast rotators.  The transition
between increasing X-ray flux with rotation and X-ray saturation is
observed at $\log(P/\tau)=0.8\pm0.1$. These results strengthen the
hypothesis that the ``saturation'' of the angular momentum loss process
depends on the stellar mass.
\end{abstract}


\section{Introduction}

In the past 10 years, numerous rotational velocity measurements have been
obtained for low-mass pre-main sequence stars and members of nearby young
open clusters.  This set of observations has shown that the angular
momentum of each star follows an evolutionary scenario from the star's
birth to the age of the Sun.  Though several models have been developed to
describe the rotational history of low-mass stars, the data available to
confront models with observations were still insufficient to provide an
unambiguous validation of these models. The measurements of stellar rotation
for various ages and the understanding of its evolution shall provide a way
to get a better knowledge of the physical processes experienced by the star
through its history. In particular, this data might offer additional insight
into the physics of young stars and their interactions with their
proto-stellar and maybe their proto-planetary disks.

On the ZAMS, G type stars ($0.8-1.0${$M_{\odot}$}) exhibit a large spread of
rotational velocities. A difference of the order of 150{\,km\,s$^{-1}$}~ is 
observed for G dwarfs in Alpha\,Per ($\approx 50$\,Myr) between
the fastest and slowest rotators (Prosser 1992). 
At Hyades age (about 600\,Myr), the spread has
disappeared and the stars with masses in the 0.6 to 1 {$M_{\odot}$}~ range show 
a monotically decreasing rotational velocity with mass (Radick et al. 1987;
Stauffer et al. 1997a).  Over this time span, the fast rotators have been
strongly braked but the slow rotators have only suffered a little braking.
Moreover, the efficiency of the braking process also depends on the stellar
mass. At the Pleiades age ($\approx100$\,Myr), almost all G type stars have
converged down to slow rotation while K dwarfs stars still exhibit a large
roational spread (Mayor \& Mermilliod 1991; Soderblom et al. 1993).  At the
Hyades age, a significant spread in rotational velocity spread is only
observed for stars less massive than 0.6{$M_{\odot}$} (Stauffer et al. 1997a).

Models have difficulties to simultaneously reproduce the large diversity of
rotators at early ages and their strong convergence in a few 100\,Myr
years. The recent discovery of a bi-modal velocity distribution for
pre-main sequence stars in Taurus and Orion (see Bouvier 1994; Choi \&
Herbst 1996) suggests that the magnetic coupling between the disk and the
star (Camenzind 1990; K\"onigl 1991) could prevent the young star from
spinning up during its PMS contraction, yielding a large spread in the
initial angular momentum distribution.  On the main sequence, the star's
rotation is mostly ruled by the angular momentum losses at the surface
through a magnetized wind (Schatzmann 1962; Weber \& Davis 1967) and by the
angular momentum transfer in the inner parts of the star (Endal \& Sofia
1978; Mac\,Gregor \& Brenner 1991).  Depending on the efficiency of the
transfer and the disk lifetime, various scenarios can be predicted. The
latest models of angular momentum evolution are from Krishnamurthi et al.
(1997a), Bouvier et al. (1997b) and Allain (1998).  

Numerous data on stellar rotation were available to these models, though
not enough yet to derive the complete distributions of rotational velocity
in young open clusters. The comparison between such distributions and the
initial velocity distribution gathered from T\,Tauri observations would
provide an important additional constraint on the models. 

Unbiased rotational velocity distributions are hard to build, numerous
observations being needed to get a statistically significant distribution.
In this case, measurements of the projected velocity ({$v\sin i$}~) are a good
alternative to the direct measurement of rotation periods which requires a
huge number of photometric measurements, difficult to achieve for a large
sample of stars. Indeed, since the pioneering work of van Leeuwen \& Alphenaar
(1982), there have been several attempts to measure photometric periods in
the Pleiades (see O'Dell et al. 1995 for a census of the observations;
Krishnamurthi et al. 1997b).  Up to now, 42 periods have been measured in
this cluster, for stars with {(B$-$V)$_{0}$} in the range 0.5 to 1.4. The 
detection of
rotational periods for the whole sample of Pleiades stars in this mass
range (more than 200 catalogued members) is difficult and will still take a
few years.
  
Measurements of {$v\sin i$}~ are more straighforward but are obviously affected
by projection effects. Yet, the knowledge of the {$v\sin i$}~ for a large
stellar sample does provide  rich {\sl statistical} information. It is
then possible to extract an accurate estimate of the distribution of {\it
equatorial} velocities in a cluster from a large number of observed
{\sl projected} rotational velocities if the stellar sample is
statistically representative of the stellar population of the cluster and
if it does not suffer from incompleteness (i.e., all the {$v\sin i$}~ must be
resolved). We consider that the inclination angles are randomly distributed in 
clusters.

In the next section we present new {$v\sin i$}~ measurements for 235 Pleiades
low-mass members.  We show that by taking into account the spectral type of
stars, we can calibrate the intrinsic width of the spectral lines and
measure the {$v\sin i$}~ of very slow rotators down to 1.5{\,km\,s$^{-1}$}. The 
calibration process is described in Sect.~2.  
We detect rotational broadening for all
observed stars. More than 98\% Pleiades members of the original Hertzprung
sample now have a resolved rotational velocity measurement up to 
{(B$-$V)$_{0}$}=1.35.
The distributions of projected rotational velocities are then converted
into distributions of equatorial velocities using a numerical algorithm
described in Sect.~3.  The velocity distributions for various mass ranges
are presented in Sect.~4. In Sect.~5 we compare our results with those
obtained for other open clusters and discuss the implications for the
models of angular momentum evolution. We also examine the relationship between X-ray flux and stellar
rotation for Pleiades stars.

\section{The {$v\sin i$}~ measurements}

The ultimate technique to measure projected equatorial rotational 
velocity ({$v\sin i$}~)
is certainly the Fourier transform technique (Smith \& Gray 1976) if the
signal--to--noise of the spectra is high.  The apparent magnitude of low mass
stars in the Pleiades is however too faint for using such a technique.
Cross-correlation algorithms which concentrate the spectral information
have to be used. They allow us to measure the mean broadening of lines with
a high precision even from low signal-to-noise spectra (Queloz 1995 and
references herein). This technique is very efficient for F, G and K stars
with slow and moderate rotation rates: applied to high resolution spectra
covering a large wavelength range, it is known to yield projected
rotational velocity with a precision of the order of 1{\,km\,s$^{-1}$}~ for slow
rotators (Benz \& Mayor 1981, 1984).

For this study we used the {{\footnotesize CORA\-VEL}}~ (Baranne et al. 1979) 
and {{\footnotesize ELO\-DIE}}~ (Baranne
et al. 1996) spectrographs. {{\footnotesize CORA\-VEL}}~ is equipped with a 
mask, located in its
focal plane. It builds cross-correlation functions by "optical" ways.
{{\footnotesize ELO\-DIE}}~ is the fiber-fed echelle spectrograph of the 
193cm-telescope of the
Haute-Provence Observatory. The echelle spectra are recorded on a 1024x1024
CCD and the cross-correlation functions are computed by an automatic
reduction carried out directly after the observations using a K0 spectral
type template (see  Baranne et al.  1996 for details about the reduction
process).

Although the way to compute  the cross-correlation functions  is
different for both instruments, the algorithm is similar. They both yield
Gaussian shaped cross-correlation functions whose interpretation is
straightforward and perfectly fitted for rotation analysis. In both cases,
the width of the cross-correlation function is estimated by fitting a
Gaussian function.  The accuracy of {{\footnotesize ELO\-DIE}}~ observations is 
better than those of {{\footnotesize CORA\-VEL}}~  due to its higher
spectral resolution but the {$v\sin i$}~
measurements of both instruments are in good agreement (see below).

The mean broadening measured by the cross-correlation technique includes
all intrinsic sources of line broadening such as micro and macroturbulence,
pressure and magnetic Zeeman splitting. Therefore, the differential
broadening of the cross-correlation function between two stars cannot
directly be interpreted as only an effect of rotational broadening.  This
is probably the main limit to the accuracy of the {$v\sin i$}~ determinations
for slow rotators. However, the calibration of the {$v\sin i$}~ carried out for
{{\footnotesize ELO\-DIE}}~ (see below) and those of Benz \& Mayor (1984) for 
{{\footnotesize CORA\-VEL}}~ shows that
a reliable interpretation of the differential broadening as a projected
rotation velocity for F, G and K dwarf stars is possible down to a value of
2{\,km\,s$^{-1}$}~ if both the spectral type and metallicity of stars are taken 
into account. The derivation of the rotational velocity distribution free of
systematic effects through a range of mass for open cluster stars is thus 
possible.

\subsection{Calibration of the {{\footnotesize ELO\-DIE}}~ {$v\sin i$}~ for 
slow rotating dwarf stars}

The effect of stellar rotation on the spectral lines can be fairly well
approximated by a convolution between a non-rotating spectrum and the
analytical broadening function described in Gray (1976). This function,
which only takes into account the geometrical effect of a
solid body rotation and the center-to-limb darkness effect, 
is accurate enough for slow rotator
measurements considering the uncertainties on the intrinsic widths of
stellar spectral lines. Compared to a complete treatment with integration
of spectra on a grid, this approximation leads to errors less than 
1{\,km\,s$^{-1}$}~(Benz \& Mayor 1981, Marcy \& Chen 1992).  
In comparison the uncertainties in the knowledge of intrinsic widths of
the cross-correlation reach
1.5{\,km\,s$^{-1}$}~ (see below).

In the slow rotator domain, up to {$v\sin i$}~ 20{\,km\,s$^{-1}$}, with our 
spectral resolution, the shape of the
cross-correl\-ation function is very well approximated by a Gaussian.
Therefore, the effect of the rotationnal broadening can be expressed as a
quadratic broadening of the cross-correlation function:
\begin{equation}
v\sin i=A\sqrt{\sigma^2-\sigma_0^2},\label{equ_rot_calcv}
\end{equation}
where $\sigma$ is the observed width, $\sigma_0$ the intrinsic width (no
rotation) and $A$ the constant coupling the differential broadening of the
cross-correlation functions to the {$v\sin i$}~ of stars. With a sample of
non-rotating stars ranging from G5 to M0 (see in
Table~\ref{table_std_calibr0} for references), 
a mean value for $A$ is measured. The
result of the computation of the $A$ constant from various non-rotating
dwarf stars artificially broadened up to 20{\,km\,s$^{-1}$}~ is listed in
Table~\ref{table_std_calibrC}. In this rotation domain no difference
greater than 0.3{\,km\,s$^{-1}$}~ has been measured  between the quadratic 
broadening model (Eq.~1) and the input rotation. A mean value of $A=1.9\pm0.1$ 
is used in this work.

\begin{table}[!htpf]
\caption[]{ $A$ constant computed  with
various non-rotating dwarf stars artificially broadened up to 
20{\,km\,s$^{-1}$}. \label{table_std_calibrC}}
\begin{flushleft}

\begin{tabular}{lll}
\noalign{\smallskip}
\hline
\noalign{\smallskip}
Name &B--V&A\\
\noalign{\smallskip}
\hline
\noalign{\smallskip}
HD\,115641&0.7&1.99\\
HD\,185144&0.8&1.86\\
HD\,10476&0.84&1.98\\
HD\,4628&0.88&1.96\\
HD\,16169&0.95&1.93\\
HD\,190007&1.15&1.85\\
HD\,201092&1.37&1.86\\
\noalign{\smallskip}
\hline
\noalign{\smallskip}
\multicolumn{2}{l}{mean value}&\multicolumn{1}{l}{1.92}\\
\noalign{\smallskip}
\hline
\end{tabular}

\end{flushleft}
\end{table}

The parameter $\sigma_0$ represents the mean intrinsic width for
non-rotating stars. It depends on the
instrumental profile and all intrinsic broadening phenomena affecting the
spectral lines. Its precise knowledge is crucial to measure slow rotators
below 5{\,km\,s$^{-1}$}~.

\subsubsection{The intrinsic width of the cross-correlation function}

\begin{figure}[!htpf]
  \psfig{width=8cm,height=8cm,file=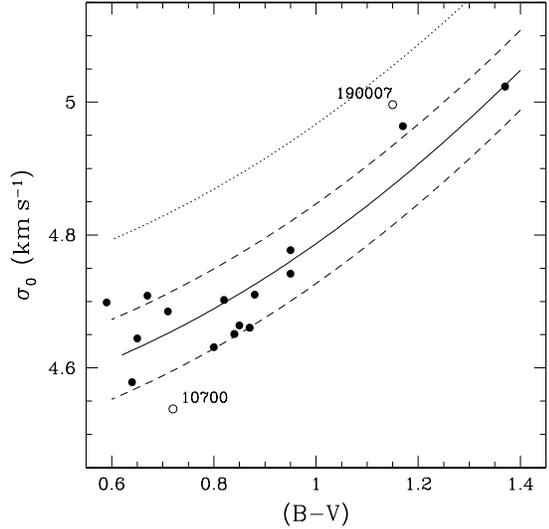}
\caption{Intrinsic width of the cross-correlation function ($\sigma_0$) vs
  $(B-V)$ for a sample of non-rotating dwarf stars. $\sigma_0$ is computed
  with Eq.~\ref{equ_rot_calcv} using column (3) and (4) of
  Table~\ref{table_std_calibr0}. The solid line is Eq.~\ref{equ_rot_calc}
  (fitted without HD\,10700 and HD190007 (open dots) values; see 
  discussion in text).  
  Dashed lines indicate a 60{\,m\,s$^{-1}$}~ uncertainty on the
  knowledge of the intrinsic width of the cross-correlation function
  corresponding to {$v\sin i$}~ $\pm1.5${\,km\,s$^{-1}$}. The dotted line 
  illustrates the $\sigma$-width of  cross-correlation functions for
  stars with a rotation broadening  
  {$v\sin i$}$=3${\,km\,s$^{-1}$}.\label{photrot2}}
\end{figure}

{{\footnotesize ELO\-DIE}}~ is an extremely stable instrument. It is kept in an 
isolated room
with a temperature control and is fed by optical fibers. The instrumental
profile is therefore very stable from one run to the other.  In order to
measure the value of the $\sigma_0$ parameter for a mass domain from
1.2\,M$_\odot$ to 0.5\,M$_\odot$, a set of non-rotating dwarf stars has
been observed ranging from G0 to M0  (Table~\ref{table_std_calibr0}).
With the reasonable assumption that all non-rotating stars with the same
temperature and the same metallicity share similar intrinsic broadenings,
this set of stars yields the following calibration (see 
Fig.~\ref{photrot2}):
\begin{equation}
\sigma_0=0.27 (\hbox{B}-\hbox{V})^2 + 4.51 \; (\pm 0.06)\; 
(\hbox{{\,km\,s$^{-1}$}~}).\label{equ_rot_calc}
\end{equation}
The 60{\,m\,s$^{-1}$}~ typical uncertainty on the individual broadening of
cross-correlation functions of each star yields a 1.5{\,km\,s$^{-1}$}~ lower 
limit for
{$v\sin i$}~ measurements by our technique. It is worth stressing that such
uncertainty has little impact on the {$v\sin i$}~ measurements of rotators
faster than 3{\,km\,s$^{-1}$} (see  Fig.~\ref{photrot2}).

This calibration is also limited to stars with solar metallicities. In
particular the intrinsic widths of cross-correlation functions for stars
HD\,10700 and HD\,190007 differ significantly from those of other stars
having a similar temperature because of their specific metallicity content.
The star HD\,10700 is a metal poor star ([Fe/H]=--0.57 from Arribas \& 
Crivellari 1989) and the star HD\,190007 belongs to the "super-metal-rich"
class stars  (Taylor\& Johnson 1987). 
All the other calibration stars have solar type
metallicities.  Since a rise of metallicity increases the number of
saturated lines, the mean spectral width, seen by the correlation process,
is slightly broadened. A diminution of the stellar metallicity content
induces the opposite effect. The use of the above calibration for deficient
and metal-rich stars would induce, respectively,  a systematic {$v\sin i$}~
underestimation and a systematic {$v\sin i$}~ over-estimation.

\begin{table}[!htpf]
\caption[]{Observed {{\footnotesize ELO\-DIE}}~ cross-correlation width for a 
set of non-rotating  stars. Column number:  (3) {$v\sin i$}~
from (a) Soderblom 1983 and (b) Gray 1984 (spectroscopic determination), (c)
Noyes et al. 1984 (estimation from $S$ index), (d) Baliunas et al. 1983
(estimation from rotation period); (4) width of the {{\footnotesize 
ELO\-DIE}}~cross-correlation function; (5) {{\footnotesize ELO\-DIE}}~ {$v\sin i$}~ 
(computed with Eq.~\ref{equ_rot_calcv} and \ref{equ_rot_calc} ). The {$v\sin 
i$}~ values less than zero are artifacts due to uncertainties in the width 
measurement. The two  stars with an asterisk which differ from the calibration 
are discussed in the text (see sec.~2.1.1).\label{table_std_calibr0}}
\begin{flushleft}
\begin{tabular}{lcccl}
\noalign{\smallskip}
\hline
\noalign{\smallskip}
Name&B--V&$ v\sin i$&$\sigma$&$v\sin i$\\
&&other&&{\scriptsize ELODIE}\\
&&{km\,s$^{-1}$}&{km\,s$^{-1}$}&{km\,s$^{-1}$}\\
(1)&(2)&(3)&(4)&\multicolumn{1}{c}{(5)}\\
\noalign{\smallskip}
\hline
\noalign{\smallskip}
HD\,109358&0.59&1.8$^a$&4.793&$2.5\pm0.5$\\
Sun&0.64&2.0        &4.698&$1.6\pm0.7$\\
HD\,32923&0.65&1.6$^a$&4.720&$1.8\pm0.6$\\
HD\,217014&0.67&1.7$^a$&4.793&$2.3\pm0.5$\\
HD\,115617&0.71&0.0$^b$    &4.685&$1.1\pm1.0$\\
HD\,10700$^\star$&0.72&0.9$^b$     &4.563&$-1.7\pm0.6$\\
HD\,185144&0.80&0.4$^b$    &4.635&$-1.3\pm0.9$\\
HD\,26965&0.82&1.0$^c$     &4.750&$1.1\pm1.0$\\
HD\,10476&0.84&0.9$^c$     &4.677&$-0.9\pm1.2$\\
HD\,3651&0.85&0.7$^d$      &4.680&$-0.9\pm1.2$\\
HD\,166620&0.87&0.8$^d$    &4.681&$-1.0\pm1.1$\\
HD\,4628&0.88&0.9$^d$      &4.735&$0.7\pm1.5$\\
HD\,16160&0.95&0.7$^d$     &4.758&$1.3\pm0.9$\\
HD\,160346&0.95&1.0$^d$    &4.806&$0.4\pm1.7$\\
HD\,190007$^\star$&1.15&1.0$^d$    &5.023&$2.4\pm0.5$\\
HD\,201091&1.17&0.8$^d$    &4.980&$1.9\pm0.6$\\
HD\,201092&1.37&0.5$^d$    &5.031&$0.7\pm1.7$\\
\noalign{\smallskip}
\hline
\end{tabular}
\end{flushleft}
\end{table}

We selected the same kind of analytical function as those used by Benz \&
Mayor (1994) to model the $\sigma_0$ change versus B--V. Such a choice
already proved to be fairly good. We decided that the
correction for high rotators included by Benz \& Mayor (1984) was not worth
adding to our {{\footnotesize ELO\-DIE}}~ $\sigma_0$ calibration considering 
the small rotational velocities of our program stars. 
The calibration found for {{\footnotesize ELO\-DIE}}~ is very
close to the {{\footnotesize CORA\-VEL}}~ one.  A rescaling of the 
{{\footnotesize ELO\-DIE}}~ result
(Eq.~\ref{equ_rot_calc}) to {{\footnotesize CORA\-VEL}}~ resolution
($\sigma^2(\hbox{{{\footnotesize CORA\-VEL}}})=\sigma^2(\hbox{ {{\footnotesize 
ELO\-DIE}}})+(5.15)^2$) leads to
$\sigma_0=6.85+0.177(\hbox{\rm B}-\hbox{\rm V})^2$, which is close to the 
direct Benz \& Mayor (1984)
calibration.

Various effects can produce the observed broadening of the
cross-correlation function with the increase of B$-$V value of stars. We
have first investigated an "instrumental effect" due to the
cross-correlation technique itself. The resolution change of {{\footnotesize 
ELO\-DIE}}~ through
the wavelength domain is too small to have any significant effect.
The cross-correlation function is the mean of spectral lines weighted by
the continuum of spectrum. Since the template (or mask) optimization
algorithm (Baranne et al. 1979) does not care about keeping a constant
resolution through the wavelength domain, the relative weighting of lines
(between blue lines and red lines) can affect the cross-correlation width.
However the observed widths of the cross-correlation functions  computed with 
the blue and the red domains of spectra do not differ significantly.

Benz \& Mayor (1984) argued that an increase of magnetic field strength in K 
and M dwarf stars is one of possible causes of the broadening of the
cross-correlation function of late type stars. Using a correlation
technique with magnetically sensitive templates (see Queloz et al. 1996 for
details) an upper limit of 1kG can be set on our late type non-rotating
standard stars. With a 1.5kG mean magnetic field upper limit, a Zeeman 
quadratic broadening of less than 1.5{\,km\,s$^{-1}$}~ is computed using a 
typical Land\'e factor
$g_{\hbox{\scriptsize eff}}=1$ for the lines.  Thus, in our case, the magnetic
field is not strong enough to be responsible of the broadening of the
cross-correlation function of our standard late type stars.

The change of damping coefficient value is probably the main cause of the
observed line broadening of late type stars. For most lines in cool stars
the pressure broadening is of Van Der Waal type. The behavior of the
$\gamma_6$ damping constant related to this broadening mechanism can be
approximated as follow: $\log(\gamma_6)\sim\log(P)-0.7\log(T)$ (Gray 1976).
Therefore, when the temperature decreases the damping parameter becomes
stronger and the spectral lines broader. Since the macroturbulence
decreases with the temperature (Gray 1984), the competition between the
pressure broadening and the macroturbulence effect could be responsible for
the flattenning of the intrinsic broadening observed around B--V$=0.6$.

\subsection{Pleiades  stars\label{section_non_member}}
 
The {$v\sin i$}~ measurements of 139 Pleiades stars with $0.5<(B-V)_0<1.4$ are
listed in Table~\ref{tableA1}. Most of these stars previously  had only
{$v\sin i$}~ upper limits (see Soderblom et al. 1993). These new data
significantly increase the number of resolved rotators. In this color
range, almost all Pleiades stars located in the center of the cluster now have
a {$v\sin i$}~ measurement. 

\begin{table*}[!htpf]
\caption{{{\footnotesize CORA\-VEL}}~ and {{\footnotesize ELO\-DIE}}~ {$v\sin 
i$}~ 
  measurements of Pleiades stars from the
  Hertzprung (1947) and Soderblom (1993) samples. Most of these stars had
  previously only {$v\sin i$}~ upper limits. SB2  indicates a double line
  spectroscopic binary where the two components are unresolved. An extra ``a''
  or ``b'' character added to the name of stars indicates each component of a
  SB2 system. The stars detected as  ``no dip'' by {{\footnotesize CORA\-VEL}}~ 
 are unresolved very 
fast rotators ({$v\sin i$}$>60${\,km\,s$^{-1}$}).}\label{tableA1}
\begin{flushleft}
\begin{tabular}{llll|llll|llll}
\hline
\noalign{\smallskip}
\multicolumn{1}{l}{star} & \multicolumn{1}{l}{\footnotesize {B--V}$_0$} & 
\multicolumn{2}{c}{{$v\sin i$}~ ({km\,s$^{-1}$})} & \multicolumn{1}{l}{star} & 
\multicolumn{1}{l}{\footnotesize {B--V}$_0$} & \multicolumn{2}{c}{{$v\sin i$}~ 
({\,km\,s$^{-1}$})} & \multicolumn{1}{l}{star} & 
\multicolumn{1}{l}{\footnotesize {B--V}$_0$} & \multicolumn{2}{c}{{$v\sin i$}~ 
({km\,s$^{-1}$})} \\
\multicolumn{1}{l}{HII} & \multicolumn{1}{l}{} & \multicolumn{1}{l}{\scriptsize 
CORAVEL} & \multicolumn{1}{l}{\scriptsize ELODIE} & \multicolumn{1}{l}{HII} & 
\multicolumn{1}{l}{} & \multicolumn{1}{l}{\scriptsize CORAVEL} & 
\multicolumn{1}{l}{\scriptsize ELODIE} & \multicolumn{1}{l}{HII} & 
\multicolumn{1}{l}{} & \multicolumn{1}{l}{\scriptsize CORAVEL} & 
\multicolumn{1}{l}{\scriptsize ELODIE}\\
\noalign{\smallskip}
\hline
25 &0.443 &44.2$\pm$4.9 & &                      793 &1.378 & &5.0$\pm$0.8&     
                      1593 &0.716 &3.0$\pm$1.4 &1.8$\pm$1.0 \\           
34 &0.888 &7.3$\pm$1.1 &5.9$\pm$0.8 &		 799 &1.287 & &4.7$\pm$0.8 &    
                      1613 &0.494 &20.1$\pm$0.8 &\\                      
97 &1.043 & &6.8$\pm$0.8&			 870 &0.68 & &9.7$\pm$0.9 &     
                      1726 &0.505 &12.9$\pm$0.7 & \\                      
102 &0.684 &18.3$\pm$0.5 & &			 879 &1.032 &7.2$\pm$1.3 
&5.6$\pm$0.8&                1756 &1.325 & &5.0$\pm$0.8 \\                      
  
120 &0.666 &9.4$\pm$1.2 & &			 882 &1.040 &no dip & &         
                      1766 &0.435 &22.7$\pm$1.3 &\\                       
129 &0.830 &5.3$\pm$1.4 &&			 883 &1.086 & &3.8$\pm$0.8 &    
                      1776 &0.680 &10.3$\pm$1.0 &9.8$\pm$0.9 \\           
152 &0.647 &11.1$\pm$1.2 &11.8$\pm$0.9 &	 885 &0.988 &6.1$\pm$1.1 
&5.2$\pm$0.8&                1785 &1.361 & &6.5$\pm$0.8 \\                      
 
164 &0.459 &38.9$\pm$5.1 & &			 890 &1.344 & &5.4$\pm$0.8 &    
                      1794 &0.589 &11.2$\pm$0.8 &\\                       
173{\footnotesize  a} &0.73 &7.8$\pm$1.2 &&	 915 &1.199 & &9.3$\pm$0.8 &    
                      1797 &0.515 &19.6$\pm$0.9 & \\                      
173{\footnotesize  b} &0.83 &6.3$\pm$2.2 & &	 916 &0.824 &6.7$\pm$1.0 
&5.6$\pm$0.8&                1856 &0.514 &15.4$\pm$0.9 & \\                     
 
174 &0.811 &no dip & &				 923 &0.579 &18.2$\pm$0.6 & &   
                      1924 &0.568 &14.2$\pm$0.7 &\\                       
186 &0.76 &11.1$\pm$0.7 &&			 974 &1.288 & &4.2$\pm$0.8 &    
                      2016 &1.181 & &9.8$\pm$0.9 \\                       
189 &1.328 & &4.7$\pm$0.8 &			 996 &0.603 &11.9$\pm$0.8 &&    
                      2027{\footnotesize  a} &0.76 &5.6$\pm$1.6 & \\      
191 &1.36 & &9.1$\pm$0.8 &			 1015 &0.609 &9.6$\pm$1.1 & &   
                      2106 &0.823 &3.7$\pm$1.5 &8.0$\pm$0.8\\             
193 &0.758 &6.3$\pm$1.4 &6.8$\pm$0.8&		 1032 &0.733 &37.2$\pm$1.9 & &  
                      2126 &0.815 &2.6$\pm$1.7 &5.3$\pm$0.8 \\            
233 &0.493 &14.1$\pm$0.7 & &			 1039 &0.958 &$<$ 5 
&4.9$\pm$0.8&                     2147{\footnotesize  a} &0.742 &6.9$\pm$3.2 & 
\\     
248 &0.738 &12.1$\pm$1.1 & &			 1061 &1.358 & &7.1$\pm$0.8 &   
                      2147{\footnotesize  b} &0.742 &10.8$\pm$2.3 &\\     
250 &0.637 &6.9$\pm$1.1 &5.9$\pm$0.8&		 1095 &0.858 &3.6$\pm$2.6 
&3.6$\pm$0.8 &              2172 &0.582 &10.3$\pm$0.4 & \\                      
253 &0.648 &38.2$\pm$1.8 & &			 1100 &1.088 & &5.4$\pm$0.8&    
                      2209 &1.336 & &5.4$\pm$0.8 \\                       
263 &0.836 &7.8$\pm$0.8 & &			 1101 &0.572 &SB2 & &           
                      2278 &0.828 &6.1$\pm$0.9 &7.7$\pm$0.8\\             
293 &0.665 &6.6$\pm$1.0 &5.1$\pm$0.8&		 1110 &1.170 & &5.9$\pm$0.8 &   
                      2284 &0.743 &3.5$\pm$1.3 &3.7$\pm$0.8 \\            
296 &0.799 &14.7$\pm$0.9 & &			 1114 &1.357 & &7.3$\pm$0.8&    
                      2311 &0.780 &6.2$\pm$1.3 &6.5$\pm$0.8 \\            
298 &0.835 &6.6$\pm$1.1 &6.3$\pm$0.8 &		 1117{\footnotesize  a} &0.62 
&6.4$\pm$2.6 & &        2341 &0.672 &3.4$\pm$1.7 &3.6$\pm$0.8\\             
303 &0.856 &17.4$\pm$0.7 &&			 1117{\footnotesize  b} &0.67 
&3.6$\pm$2.5 & &        2366 &0.779 &2.9$\pm$1.7 &4.3$\pm$0.8 \\            
314 &0.613 &41.9$\pm$1.6 & &			 1122 &0.425 &28.6$\pm$1.2 &&   
                      2406 &0.720 &9.2$\pm$0.5 &8.5$\pm$0.8 \\            
320 &0.840 &10.8$\pm$0.5 & &			 1124 &0.924 &5.6$\pm$3.3 
&3.5$\pm$0.8 &              2407$^\dagger$ &0.910 & &6.3$\pm$0.8\\              
338 &0.428 &$>$40 &&				 1132 &0.451 &$>$40 & &         
                      2462 &0.791 &5.4$\pm$1.2 &4.9$\pm$0.8 \\            
345 &0.806 &18.9$\pm$0.7 & &			 1136 &0.784 &no dip &&         
                      2500 &0.58 &33.0$\pm$3.4 & \\                       
357 &1.182 & &10.0$\pm$0.9 &			 1139 &0.437 &31.4$\pm$1.9 & &  
                      2506 &0.552 &13.8$\pm$0.8 &\\                       
380 &1.171 & &6.0$\pm$0.8&			 1182 &0.597 &16.4$\pm$1.1 & &  
                      2548 &1.297 & &5.7$\pm$0.8 \\                       
405 &0.495 &18.2$\pm$1.0 & &			 1200 &0.507 &13.7$\pm$0.9 &&   
                      2588 &1.133 & &5.2$\pm$0.8 \\                       
430 &0.777 &7.3$\pm$1.0 &6.3$\pm$0.8 &		 1207 &0.591 &5.1$\pm$1.3 & &   
                      2644 &0.698 &3.6$\pm$1.9 &5.0$\pm$0.8\\             
451 &1.169 & &5.7$\pm$0.8&			 1215 &0.599 &6.5$\pm$0.9 
&4.1$\pm$0.8 &              2665 &0.790 &5.9$\pm$1.2 &5.4$\pm$0.8 \\            
476 &0.53 &21.0$\pm$1.0 & &			 1220 &0.830 &4.8$\pm$1.3 
&6.3$\pm$0.8&               2741 &0.972 & &7.8$\pm$0.8 \\                       
489 &0.594 &18.3$\pm$0.8 & &			 1275 &0.791 &6.4$\pm$1.2 
&6.4$\pm$0.8 &              2786 &0.562 &22.0$\pm$1.0 &\\                       
513 &1.275 & &7.4$\pm$0.8&			 1298 &0.957 &5.6$\pm$2.2 
&4.8$\pm$0.8 &              2870 &0.974 & &4.0$\pm$0.8 \\                       
514 &0.657 &10.5$\pm$1.0 & &			 1332 &0.988 &$<$ 2 
&5.3$\pm$0.8&                     2880 &0.818 &6.3$\pm$1.1 &6.0$\pm$0.8 \\      
      
522 &0.879 &3.6$\pm$0.8 &4.4$\pm$0.8 &		 1348{\footnotesize  a} &1.05 & 
&5.1$\pm$0.8 &        2881$^\star$ &0.920 &7.8$\pm$1.1 &13.3$\pm$1.0\\    
571 &0.748 &7.6$\pm$0.5 &6.8$\pm$0.8&		 1348{\footnotesize  b} &1.35 & 
&1.8$\pm$0.8 &        2984 &0.961 & &5.4$\pm$0.8 \\                       
590 &1.323 & &6.9$\pm$0.8 &			 1355 &1.362 & &12.5$\pm$0.9&   
                      3019 &1.174 & &6.0$\pm$0.8 \\                       
627 &0.473 &33.2$\pm$1.3 & &			 1392 &0.55 &15.7$\pm$1.6 & &   
                      3096 &0.948 &6.2$\pm$1.3 &\\                        
636 &0.981 & &3.5$\pm$0.8&			 1454 &1.068 & &3.3$\pm$0.8 &   
                      3097 &0.697 &14.6$\pm$0.3 & \\                      
659 &0.66 &12.2$\pm$1.1 & &			 1485 &1.309 & &42.0$\pm$2.2&   
                      3104 &1.24 & &7.1$\pm$0.8 \\                        
676 &1.065 & &5.5$\pm$0.8 &			 1512 &1.207 & &5.3$\pm$0.8 &   
                      3179 &0.529 &5.5$\pm$1.2 &4.3$\pm$0.8\\             
727 &0.519 &$>$40 &&				 1514 &0.609 &13.6$\pm$0.8 & &  
                      3187 &1.139 & &6.2$\pm$0.8 \\                    
739 &0.586 &14.4$\pm$0.6 & &			 1516 &1.267 & &105$\pm$10&     
                &&&\\
746 &0.768 &4.9$\pm$1.0 &4.8$\pm$0.8 &		 1553 &1.078 & &9.6$\pm$0.9 &   
                &&&\\
\hline
\end{tabular}

\end{flushleft}
($^\dagger$) Binary systems with rotations synchronized to the orbital 
motion.\\ ($^\star$) Suspected long period  double line spectroscopic binary 
(unresolved). 
\end{table*}

In Table~\ref{tableB1}, {{\footnotesize CORA\-VEL}}~ measurements are listed 
for 82 corona
stars recently selected as new Pleiades members by Rosvick et al (1992)
and Mermilliod et al. (1997). This selection is based on proper motion,
photometry and radial velocity measurements.  For the corona stars
selected on the basis of the van Leeuwen (1983) proper motion
measurements, the probability to catch a non-member with a photometry and
a proper motion in agreement with the Pleiades ones is 25\% in a velocity
range of 60{\,km\,s$^{-1}$} around the Pleiades radial velocity (see the radial
velocity distribution of non-members compared to that of members in
Fig.~1 of Rosvick et al. 1992). For the selection of Mermilliod et al.
(1997), based on Artjukhina et al. (1970) proper motions, we expect more
contaminants (2-3 times greater) because the accuracy of the astrometry
is not as good as that of van Leeuwen (1983). Since candidate members
were selected over a radial velocity domain of $\pm2${\,km\,s$^{-1}$} (Rosvick 
et al. 1992) and $\pm4${\,km\,s$^{-1}$} 
(Mermilliod et al. 1997) around the cluster's mean
radial velocity, and considering that the radial velocity of field stars
is almost random around the cluster velocity, we may have 
perhaps 3--4 non-members hidden in our corona sample.  Such a small number
of contaminants does not have  significant impact on the final rotation
velocity distribution.

\begin{table*}[!htpf]
\caption{{{\footnotesize CORA\-VEL}}~ {$v\sin i$}~ measurement of  stars 
located in the corona of the Pleiades and recently selected by Rosvick et al. 
(1992) and by Mermilliod et al. (1997) as cluster members.}\label{tableB1}
\begin{flushleft}
\begin{tabular}{lll|lll|lll}
\hline
\noalign{\smallskip}
\multicolumn{1}{l}{star} & \multicolumn{1}{l}{{(B$-$V)$_{0}$}} & 
\multicolumn{1}{l}{ {$v\sin i$}~({\,km\,s$^{-1}$})} & \multicolumn{1}{l}{star} 
& \multicolumn{1}{l}{{(B$-$V)$_{0}$}} & \multicolumn{1}{l}{ {$v\sin 
i$}~({\,km\,s$^{-1}$})} & \multicolumn{1}{l}{star} & 
\multicolumn{1}{l}{{(B$-$V)$_{0}$}} & \multicolumn{1}{l}{ {$v\sin 
i$}~({\,km\,s$^{-1}$})}\\
\noalign{\smallskip}
\hline
R60 &0.440 &no dip &               Pels 39$^\star$ &0.83 &2.0$\pm$2.0 &  Pels 
143 &0.85 &5.2$\pm$1.2   \\	
S151x &0.46 &21.6$\pm$1.0 &	   Pels 40 &0.53 &11.9$\pm$0.7&          Pels 
146 &0.62 &17.6$\pm$1.4  \\	
S183x &0.5 &32.3$\pm$1.4&	   Pels 44 &0.86 &3.9$\pm$3.1 &          Pels 
150 &0.48 &26.3$\pm$1.7  \\	
S37 &0.403 &no dip &		   Pels 45 &0.78 &7.8$\pm$1.3 &          Pels 
151 &0.67 &5.5$\pm$2.2   \\	
S39 &0.424 &no dip &		   Pels 46 &0.7 &8.1$\pm$1.2&            Pels 
173 &0.41 &36.7$\pm$3.0  \\	
Pels 3 &0.47 &SB2&		   Pels 47 &0.86 &4.5$\pm$1.5 &          Pels 
174 &0.54 &$>$40         \\	
Pels 4 &0.78 &2.6$\pm$1.7 &	   Pels 50 &0.8 &7.0$\pm$1.5 &           Ia 317 
&0.5 &$>$40            \\	
Pels 5 &0.66 &11.1$\pm$0.9 &	   Pels 56 &0.77 &$>$40&                 Ib 038 
&0.66 &10.9$\pm$0.9    \\	
Pels 6 &0.47 &35.9$\pm$3.1&	   Pels 60 &0.52 &no dip &               Ib 055 
&0.74 &6.6$\pm$1.0     \\	
Pels 7 &0.62 &2.7$\pm$1.9 &	   Pels 68 &0.81 &6.7$\pm$2.0 &          Ib 078 
&1.03 &6.4$\pm$1.1     \\	
Pels 8 &0.64 &15.5$\pm$1.0 &	   Pels 69 &0.81 &16.3$\pm$0.8&          Ib 
146a &0.45 &12.4$\pm$3.5   \\	
Pels 11 &0.82 &5.2$\pm$1.3&	   Pels 71 &0.77 &9.1$\pm$1.1 &          Ib 
146b &0.5 &8.6$\pm$5.3     \\	
Pels 12 &0.69 &11.2$\pm$0.9 &	   Pels 75 &0.87 &$>$40 &                Ib 288 
&1.01 &SB2             \\	
Pels 15 &0.54 &24.4$\pm$1.1 &	   Pels 78 &0.7 &8.8$\pm$1.1&            II 293 
&0.65 &8.5$\pm$0.9     \\	
Pels 17 &0.5 &29.3$\pm$2.0&	   Pels 79 &0.73 &2.8$\pm$1.7 &          II 359 
&0.57 &16.6$\pm$0.9    \\	
Pels 18 &0.58 &11.8$\pm$1.0 &	   Pels 83 &0.69 &14.5$\pm$1.1 &         III 
059 &0.8 &4.1$\pm$1.2     \\	
Pels 19 &0.85 &4.8$\pm$1.4 &	   Pels 86 &0.45 &no dip&                III 
079 &0.43 &no dip         \\	
Pels 20 &0.61 &9.6$\pm$1.1&	   Pels 89 &0.84 &3.2$\pm$2.1 &          III 
391 &0.73 &2.3$\pm$1.8    \\	
Pels 23 &0.55 &36.2$\pm$1.7 &	   Pels 121 &0.62 &4.9$\pm$1.1 &         III 
679 &0.96 &4.1$\pm$1.5    \\	
Pels 25 &0.443 &no dip &	   Pels 123 &0.92 &3.1$\pm$2.1&          III 
700 &0.67 &9.3$\pm$0.9    \\	
Pels 28 &0.94 &8.9$\pm$1.2&	   Pels 124 &0.5 &20.5$\pm$0.9 &         IV 131 
&0.77 &2.2$\pm$1.5     \\	
Pels 29 &0.6 &33.8$\pm$1.9 &	   Pels 126 &0.58 &9.8$\pm$1.0 &         IV 314 
&0.98 &2.7$\pm$2.8     \\	
Pels 30 &0.9 &6.7$\pm$1.1 &	   Pels 128 &0.61 &4.6$\pm$1.5&          V 088 
&0.77 &5.2$\pm$1.2      \\	
Pels 31 &0.94 &12.2$\pm$1.0&	   Pels 135 &0.45 &no dip &              V 198a 
&0.6 &10.5$\pm$1.1     \\	
Pels 34 &0.69 &3.6$\pm$1.4 &	   Pels 137 &1.01 &3.6$\pm$1.6 &         V 198b 
&0.87 &2.0$\pm$2.0     \\	
Pels 35 &0.51 &19.4$\pm$1.1 &	   Pels 138 &0.76 &5.6$\pm$1.6&          V 308 
&0.78 &8.0$\pm$1.0      \\	
Pels 37 &0.63 &13.4$\pm$2.6&	   Pels 140 &0.45 &11.5$\pm$1.2 &        && \\
Pels 38 &0.73 &3.1$\pm$2.0 &	   Pels 142 &0.81 &3.9$\pm$1.5 &         &&\\
\hline
\end{tabular}
\end{flushleft}
($^\star$) Pels\,39 lies slightly  above the main sequence but doesn't show
any hint of binarity with radial velocities. However, its very small {$v\sin 
i$}~ value is  an  indication that this star could be a non-member.
\end{table*}

All the  {$v\sin i$} measurements done with {{\footnotesize CORA\-VEL}}~ are 
based on many 
observations spread over the past 15 years. In order to avoid any statistical 
bias in the measurement of the 
mean {$v\sin i$} per star (Eq.~1 is highly non-symmetric when $\sigma$ is close 
to $\sigma_0$), 
a weighted mean  broadening ($\sigma$) was first computed for each star and 
then transformed into {$v\sin i$}  according to the Benz \& Mayor (1981, 1984) 
calibration.  
To take into  account the instrumental uncertainties due to the 
variation of the point spread function of {{\footnotesize CORA\-VEL}}~ from 
night to night and be consistent with the dispersion measured on the  mean 
broadening ($\sigma$) of the cross-correlation function of each star, an 
internal error of $0.2${\,km\,s$^{-1}$}  has been added  to the photon noise of 
each measurements.   Most of  the  {{\footnotesize ELO\-DIE}}~ {$v\sin i$} 
are computed from only  one spectrum. 
When two measurements were available we used the same procedure that for 
{{\footnotesize CORA\-VEL}}~ measurements but without  any extra intrumental 
uncertainty since the stability of the  mean width of point 
spread function of {{\footnotesize ELO\-DIE}} proved to be better than 
$\pm10${\,m\,s$^{-1}$}. Both {{\footnotesize ELO\-DIE}}~ 
and {{\footnotesize CORA\-VEL}}~ {$v\sin i$} error bar estimates listed 
in this paper are in agreement with  the internal errors displayed by the 
dispersion of individual measurements of each star.

\begin{figure}[!htpf]
  \psfig{width=8cm,height=8cm,file=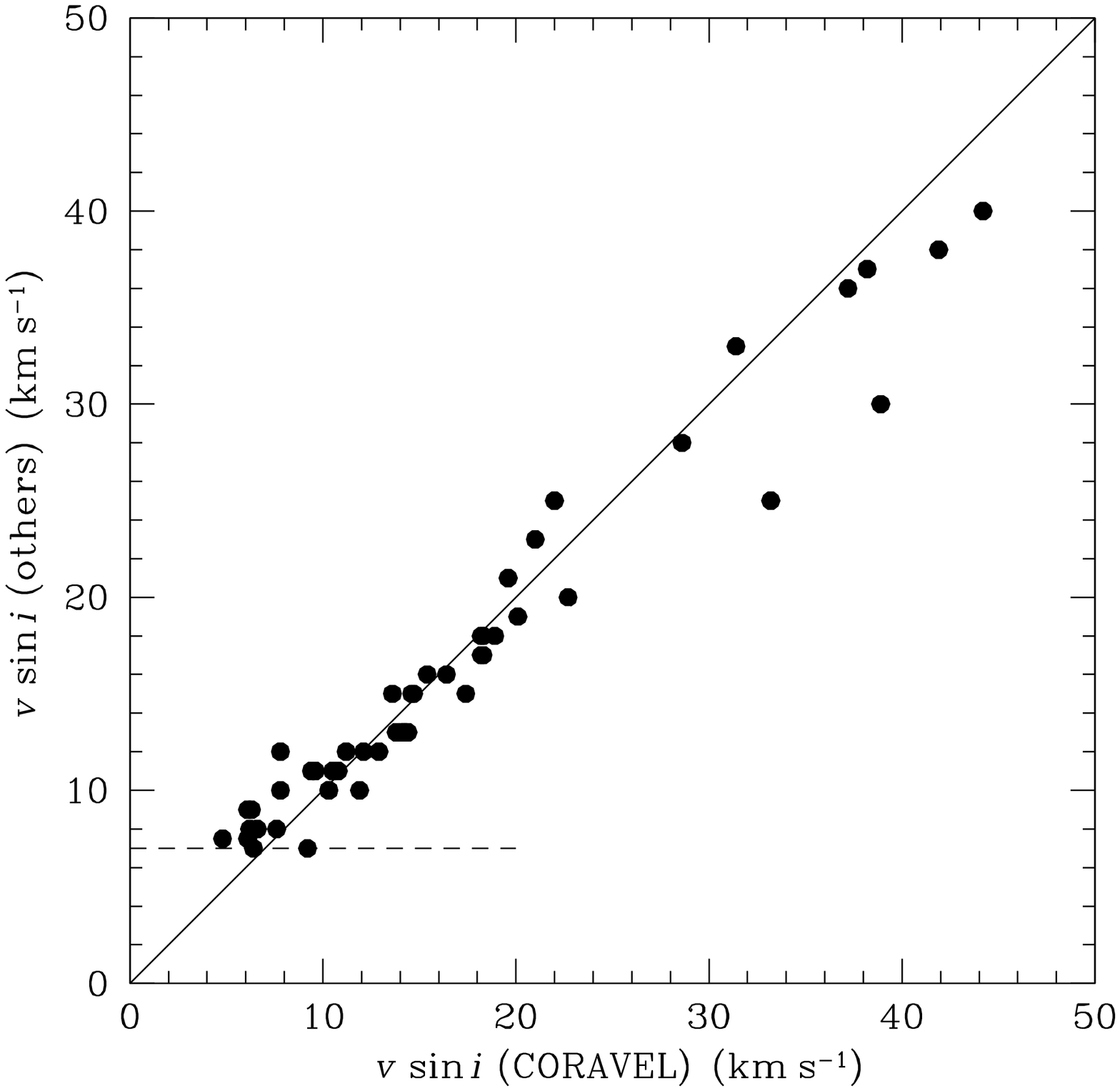}
\caption{  
  The comparison between {$v\sin i$}~ measurements from {{\footnotesize 
  CORA\-VEL}}~ and from
  Soderblom et al.  (1993) is displayed for common Pleiades stars. The dashed line
  indicates the lower limit for Soderblom et al. (1993) measurements. For
  rotators faster than 30{\,km\,s$^{-1}$}, the {{\footnotesize CORA\-VEL}}~ 
  measurements exhibit a slight
  deviation from the one-to-one relation displayed by the solid line. This
  effect is caused by the limited scanning domain of
  {{\footnotesize CORA\-VEL}}.}\label{corcompgs}
\end{figure}

The limited scanning domain of {{\footnotesize CORA\-VEL}}~ restricts the 
{$v\sin i$}~ measurements to
small and moderate rotators ($<40${\,km\,s$^{-1}$}). The faster rotators only 
have {$v\sin i$}~ {\it lower\/} limits.  For very fast rotating stars 
($>60${\,km\,s$^{-1}$}~),
the cross-correlation function in the scanning domain is flat. The 
{{\footnotesize CORA\-VEL}}~
measurements of stars cooler than F2 spectral type and without dip
detection (referred as "no dip" in the tables) corresponds to the
detection of fast rotating stars.  The limitation of the width of the
{{\footnotesize CORA\-VEL}}~ scanning domain leads also to a slight 
over-estimation of  the {$v\sin i$}~ for
rotators faster than 30{\,km\,s$^{-1}$}~(see Fig.~\ref{corcompgs}). In that 
case, the width of the scanning window is too small to accurately measure the
location of the continuum around the dip, and the Gaussian $\chi^2$--fit is
biased towards broader Gaussians. The {$v\sin i$}~ measurement of fast rotators
observed with {{\footnotesize ELO\-DIE}}~ has been computed by searching for 
the best match between
the observed cross-correlation functions and others computed from a
non-rotating star artificially  broadened by the Gray (1976) analytical
function.

In Table~\ref{tableC1}, {{\footnotesize ELO\-DIE}}~ measurements of 16 corona 
stars from Pels's
list suspected to be members by van Leeuwen (1983) are displayed.  All of these
stars were selected as Pleiades members based on their photometry and
proper motion. For all of them, we measure a radial velocity very close to
the mean cluster velocity. By considering the size of the radial velocity
selection window ($\pm1${\,km\,s$^{-1}$}) and by following the same argument as 
above regarding the contamination by field stars, 
we do not expect any non-members to  be hidden in this sample .

\begin{table*}[!htpf]
\caption{
{{\footnotesize ELO\-DIE}}~ {$v\sin i$}~ and radial velocity ($V_r$) 
measurements of new members of the  Pleiades corona. 
($O-C$) is the velocity difference with the mean velocity of the cluster
corrected from  the projection effect. Note that, in order to use the
convergent point measured by Rosvick et al. (1992)  to compute the ($O-C$),
the {{\footnotesize ELO\-DIE}}~ measurements have been converted into the 
{{\footnotesize CORA\-VEL}}~ zero point
reference frame.  The following velocity offset has been used:
$V_E-V_C=0.48-1.45(B-V)+0.53(B-V)^2$ (Udry et al. 1998).}
\label{tableC1}
\begin{flushleft}
\begin{tabular}{llllr|llllr}
\hline
\noalign{\smallskip}
\multicolumn{1}{l}{star} & \multicolumn{1}{l}{{(B$-$V)$_{0}$}} & 
\multicolumn{1}{l}{ {$v\sin i$}} & \multicolumn{1}{l}{$V_r$} & 
\multicolumn{1}{l}{($O-C$)}&\multicolumn{1}{l}{star} & 
\multicolumn{1}{l}{{(B$-$V)$_{0}$}} & \multicolumn{1}{l}{ {$v\sin i$}} & 
\multicolumn{1}{l}{$V_r$} & \multicolumn{1}{l}{($O-C$)}\\
\noalign{\smallskip}
\hline
Pels 9 &1.0 &5.5$\pm$0.8 &6.41 &$-0.15$ &   Pels 72 &0.72 &160$\pm$10  &6.0: 
&0.78    \\     
Pels 22 &0.90 &12.1$\pm$0.9 &5.08 &0.34&    Pels 109 &1.31 &5.5$\pm$0.8 &4.96 
&$-0.35$ \\     
Pels 41 &0.90 &5.1$\pm$0.8 &5.59 &0.56 &    Pels 114 &1.04 &6.0$\pm$0.8 &6.70 
&0.21    \\     
Pels 43 &0.92 &6.8$\pm$0.8 &4.92 &0.04&	    Pels 115 &1.06 &5.5$\pm$0.8 &6.28 
&0.24    \\     
Pels 59 &0.99 &5.6$\pm$0.8 &5.06 &0.02 &    Pels 137 &1.01 &5.2$\pm$0.8 &4.90 
&0.12    \\     
Pels 63 &1.07 &2.9$\pm$0.8 &7.41 &$-0.36$&  Pels 162 &0.91 &3.2$\pm$0.8 &6.10 
&$-0.18$ \\     
Pels 66 &1.0 &4.9$\pm$0.8 &7.39 &0.45 &	    Pels 189 &0.93 &4.9$\pm$0.8 &5.87 
&0.53    \\     
Pels 71 &0.77 &11.1$\pm$0.9 &6.25 &0.23&    Pels 192 &1.34 &11.4$\pm$0.9 &6.77 
&0.34   \\     
\hline
\end{tabular}

\end{flushleft}
\end{table*}
  
Some stars in our sample ( HII\,173, HII\,1101, HII\,1117, HII\,1348, HII\,2027,
HII\,2147,  Pels\,3, Ib\,146, Ib\,288,  V\,198) have been detected
as double-lined spectroscopic binaries (SB2). For most of them we clearly
resolved each component and we are able to measure their rotational
broadening. Moreover, since we know their mass ratio either from the
relative contrast of their cross--correlation function or from the orbit
solution (when there is one), we can also estimated their {(B$-$V)$_{0}$}~ and 
used that value to compute their {$v\sin i$}.  Each of these measurements are
identifed in Tables~\ref{tableA1} and \ref{tableB1} with the standard
notation used by spectroscopists for SB2 binaries. An extra ``a'' or ``b''
character has been added to the star name.

\begin{figure}[!htpf]
  \psfig{width=8cm,height=8cm,file=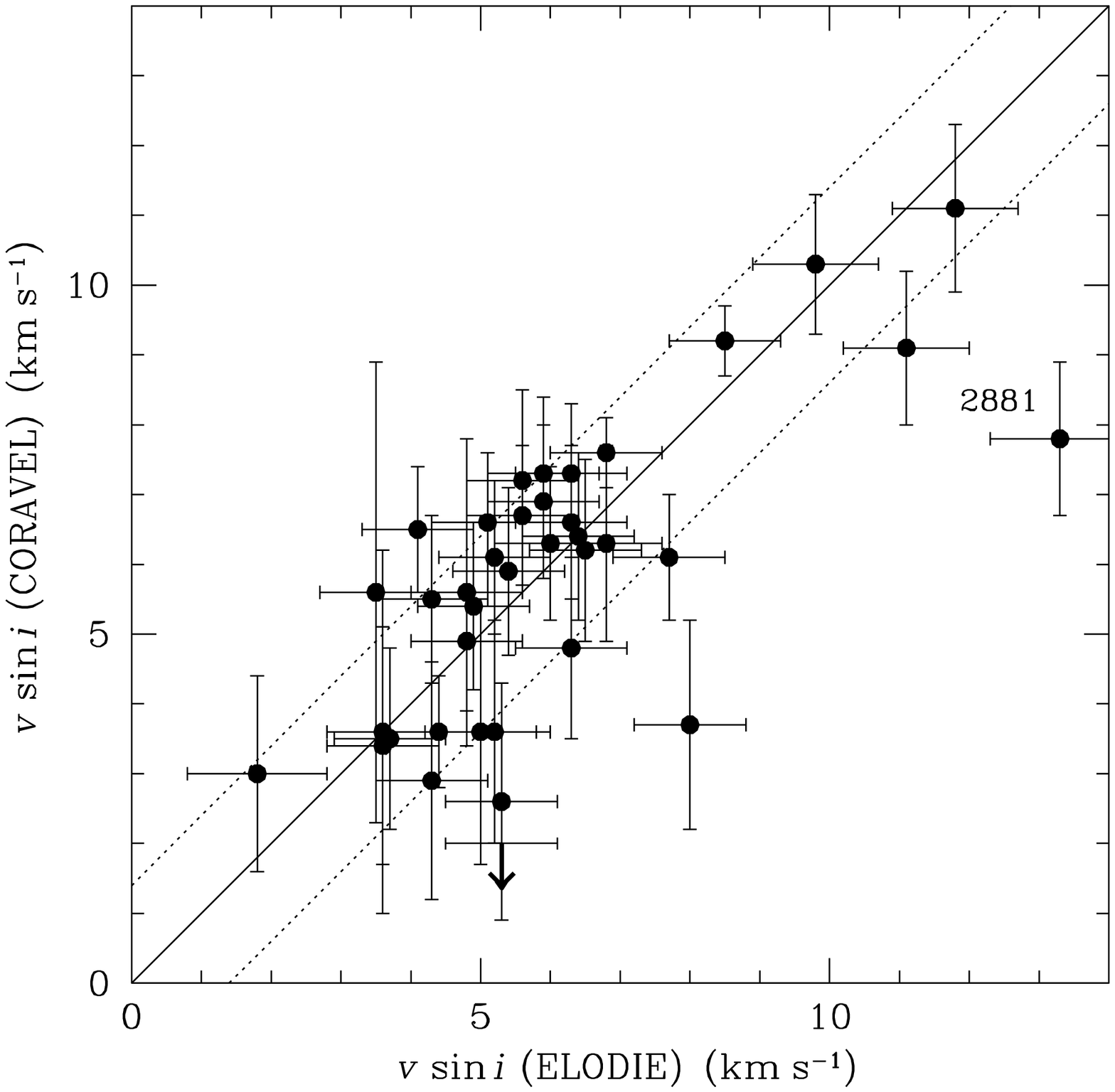}
\caption{
  Comparison between {{\footnotesize ELO\-DIE}}~ and {{\footnotesize 
  CORA\-VEL}}~ {$v\sin i$}~ measurements of Pleiades
  stars. The solid line indicates the one-to-one relation and the dotted
  lines $\pm$1.4{\,km\,s$^{-1}$}. Both data sets show a very good agreement 
  excepted
  for HII\,2881. However this star is suspected to be an unresolved SB2. It
  is a close visual binary (Bouvier et al. 1997a) and the {{\footnotesize 
  CORA\-VEL}}~ data show a radial
  velocity variability and changes of the width of the cross-correlation
  function over few years.}\label{corcompec}
\end{figure}

For all stars in common between {{\footnotesize CORA\-VEL}}~ and 
{{\footnotesize ELO\-DIE}}~ observed samples, the
comparison of the {$v\sin i$}~ measurements shows a good agreement (see
Fig.~\ref{corcompec}) with a difference less than 1{\,km\,s$^{-1}$} down to 
{$v\sin i$}~
3{\,km\,s$^{-1}$}.  The standard deviation of the {$v\sin i$}~ difference 
1.4{\,km\,s$^{-1}$}~ is in
agreement with the combined error estimate. This strengthens the
reliability of our results and clearly pushes our {$v\sin i$}~ detection limit
down to 3{\,km\,s$^{-1}$}.  Since both sets of data are coherent, they have 
been merged
together. When two measurements are available for the same star, a weighted 
mean is computed.

\begin{figure*}[!htpf]
  \psfig{width=18cm,height=13cm,file=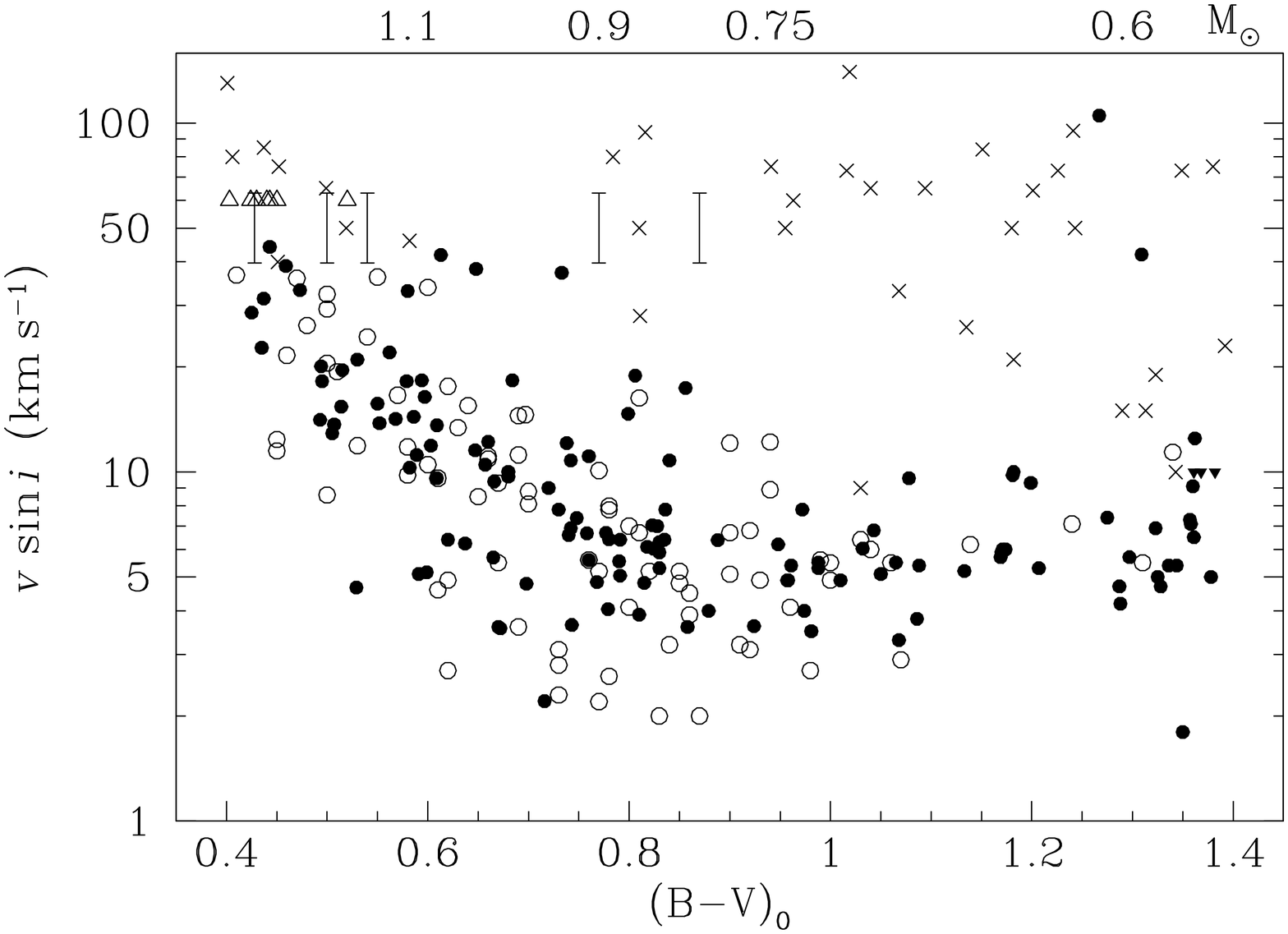}
  \caption{
    {$v\sin i$}~ of Pleiades stars  with spectral type ranging
    from F5 to M0. Measurements from this work are displayed as circles.
    The filled circles represent stars within a distance of $80'$ from the
    cluster's center and the open circles are stars in the Pleiades corona
    ($r>80'$).  The crosses represent measurements from other sources (see
    Soderblom 1993).  The triangles indicate unresolved {$v\sin i$}~ (head-up:
    lower limit, head-down: upper limit). The error bars are used for
    {{\footnotesize CORA\-VEL}}~ measurements with large {$v\sin i$}~ where 
systematic effects
    decrease the accuracy of the measurements (see text for explanations).
    The binary systems which are synchronized by tidal effects are not
    displayed.}\label{vsinibv0}
\end{figure*}

The resulting {$v\sin i$}~ distribution of low-mass Pleiades stars is shown in
Fig.~\ref{vsinibv0}. The {{\footnotesize ELO\-DIE}}~ and {{\footnotesize 
CORA\-VEL}}~ samples were supplemented with
published {$v\sin i$}~ measurements (Soderblom et al. 1993) when the star was 
not observed by {{\footnotesize ELO\-DIE}}~ or {{\footnotesize CORA\-VEL}}~ 
or when it was detected as a fast rotator with only a
{{\footnotesize CORA\-VEL}}~ lower limit.  The  data displayed in 
Fig.~\ref{vsinibv0}
represents the {$v\sin i$}~ sample used in the data analysis below to compute 
the distributions of equatorial velocity in Pleiades. Note that the following
Pleiades stars are excluded for the data analysis: HII\,761,
HII\,1338, HII\,2407.  They are all short period binaries with rotation rates
synchronized with their orbital motions (Mermilliod et al. 1992; Raboud \& 
Mermilliod 1998). Their rotation rate is driven by
the tidal interaction between each components instead of by the angular
momentum history of each star itself.

The comparison between {$v\sin i$}~ measurements and period measurements
provides an external check on our calibration.  There are 19 slow rotators
({$v\sin i$}$<15${\,km\,s$^{-1}$}) in our sample with known rotational 
periods (Prosser et al.  1995; O'Dell et al. 1995; Krishnamurthi et al. 1997b). 
We find that
the resulting {$\sin i$}~ distribution ($v\sin i/v$) is consistent with a random
{$\sin i$}~ distribution. 

We may expect the turbulence in the  atmosphere of  the young stars
to be  larger, than for the old ones -- the calibrators --, 
despite that there is no prove of it but only arguments related to a possible
extra  stirring-up in the photosphere of chromospherically active stars (spots). 
This effect would make us to over-estimate all  {$v\sin i$} measurements of 
slow rotators. However, since  nothing significant 
has been detected by the comparison with period measurements,  such effect, if 
real, has to  be very small. We are aware that such a small comparison sample
still leaves large statistical uncertainties but any systematic effects what 
would systematically change
the {$v\sin i$}~ value of slow rotators  by more than 1{\,km\,s$^{-1}$} can be 
ruled out.

The {$v\sin i$}~ distribution of the corona stars compared to that located 
in the center 
does not show any significant differences excepted perhaps in the
$0.77-0.94$ {(B$-$V)$_{0}$}~ range where a slight excess of slow rotators might 
be seen in the corona (see on Fig.~\ref{vsinibv0}). In this {(B$-$V)$_{0}$}~ 
range, the
K-S statistical test indicates a probability of 96\% that the difference
between the {$v\sin i$}~ distributions of the corona and the core of cluster
is real. It may thus be that some of the slowest rotators 
in this {(B$-$V)$_{0}$}~ range are not
Pleiades members since we do expect a few field stars to contaminate our
corona sample (see above). At this point, this difference cannot be
interpreted as an intrinsic difference between the rotational velocity
distributions of corona and core stars.

Finally, for the sake of completness, Table~\ref{tableD} lists stars
identified in the literature as probable Pleiades member but whose radial
velocity is found here to conflict with membership.  We consider these
stars as non-members. For the 3 Pels objects, this is a somewhat
conservative conclusion since we only have one velocity measurement. A
second measurement at least is necessary to exclude SB1 and draw conclusions on
their membership. However, the very small {$v\sin i$}~ values of these 3 stars
is an extra indication that they are likely to be field stars.

\begin{table}[!htpf]
\caption{New stars detected as non-members of Pleiades.\label{tableD}}
\begin{flushleft}
\begin{tabular}{llll}
\hline
\noalign{\smallskip}
\multicolumn{1}{l}{star}&\multicolumn{1}{l}{$V_r$ 
(km\,s$^{-1}$)}&\multicolumn{1}{l}{JDB}&
\multicolumn{1}{l}{$v\sin i$ (km\,s$^{-1}$)}\\
\noalign{\smallskip}
\hline
\noalign{\smallskip}
Pels 93	&   -31.2& 50394.7&  $<2$\\
Pels 117&    11.1& 50394.4&    2\\
Pels 132&    24.9& 50394.6&  $<2$\\
HII3030	&     -.7& 49752.4&  $<2$\\ 
        &     -.4& 50394.5&  $<2$\\
\noalign{\smallskip}
\hline
\noalign{\smallskip}
\end{tabular}
\end{flushleft}
\end{table}

\section{Rotational velocity distribution}

A numerical method is used to derive the distribution of equatorial
velocities from the complete sample of {$v\sin i$}. We use the statistical
relationships between the projected rotations and the equatorial rotations
derived by Chandrasekhar \& M\"unch (1950), and the numerical inversion
procedure developed by Gaig\'e (1992).

\subsection{Procedure description}

Chandrasekhar and M\"unch (1950) first expressed the analytical
relationships between the distributions of projected and equatorial
velocities, under the assumption that rotational axes are randomly
distributed in space. With $y$ = {$v\sin i$}~, its distribution $\phi(y)$ is
related to the distribution of equatorial velocities $f(v)$ by: 

\begin{equation}
 \phi(y) = y \int_{y}^{\infty} \frac{f(v)}{v(v^{2}-y^{2})^{1/2}}dv,
\end{equation}
and
\begin{equation}
f(v)=-\frac{2}{\pi} v^{2} \frac{\partial}{\partial v} v \int_{v}^{\infty}
\frac{\phi(y)}{y^{2}(y^{2}-v^{2})^{1/2}} dy.\label{fveq}
\end{equation}

In order to solve the inverse problem, i.e., to compute $f(v)$ from
$\phi(y)$ from Eq.~\ref{fveq}, the (non continuous) distribution of
projected velocities has to be differentiated. Chandrasekhar and M\"unch
suggested that it would be easier to first assume a parametric form for
$f(v)$, then to compute the corresponding {$v\sin i$}~ distribution (direct
problem) and finally adjust a set of parameters to reproduce the
observations. But this method requires assumptions on the a priori unknown
shape of $f(v)$.  The large number of {$v\sin i$}~ resolved in the Pleiades
allow us instead to use a numerical inversion method to measure $f(v)$
from $\phi(y)$.

The numerical computation of Eq.~\ref{fveq} requires that $\phi(y)$ be a
continuous fonction.  In order to transform the observed distribution into
a continuous function, we suppose that the contribution of each {$v\sin i$}~
measurement ($y_{i}$) to the final distribution $\phi(y)$ can be modeled by
a Gaussian-shape distribution probability $p_{i}(y)= 1/(l \sqrt{2\pi})
\exp(-(y-y_{i})^{2}/(2l^{2}))$, centered in $y_{i}$ and with $l$ the
$\sigma$--width of the Gaussian (Gaig\'e 1992). In other words, each
measurement is ``distributed'' over a {$v\sin i$}~ range with a Gaussian
probability. Under this assumption, the distribution of projected
velocities becomes a continuous function formed by a sum of Gaussians:
\begin{equation}
 \phi(y) = \frac{1}{N}\sum_{i=1}^{N} p_{i}(y),
\end{equation}
with $N$ the number of measurements. With these assumptions,
Eq.~\ref{fveq} is finally solved using standard differentiation and
integration algorithms (see Gaig\'e 1992 for details).

In addition, we define the cumulative distributions both for projected and
equatorial velocities:
\begin{eqnarray*}
\Phi(y)=\int_{0}^{y} \phi(x) dx&\hbox{ , }&
F(v) = \int_{0}^{v} f(w) dw\\
\end{eqnarray*}

\subsection{Tests on the stability on the inversion procedure.}

The detailed shape of the inverse distributions $f(v)$ and $F(v)$ depends
upon the number of stars $N$, and on the width of the Gaussian $l$
used to transform the observed projected distribution into a continuous one.  
In order
to test the stability of the inversion procedure and to investigate the
influence of $l$ and $N$, we  have run Monte-Carlo simulations starting
from a
synthetic initial equatorial velocity distribution  $f_0(v)$. This
distribution has been chosen so as to mimic the observed distribution in
the Pleiades: a peak at low velocity and a high-velocity tail.  The
steps of our 
Monte-carlo simulation were the following: First we 
generated a sample of $N$ equatorial velocities  $v_i$    distributed
along $f_0(v)$. Second, a random {$\sin i$}~  was applied to each velocity 
to simulate an observed   projected rotational velocity $y_{i}$. Third we
computed the continue velocity distribution $\phi(y)$. Finally we solved 
Eq.~\ref{fveq}  and
we got  a measurement of the equatorial velocity distribution $f(v)$.

\begin{figure}[!htpf]
  \psfig{width=9cm,height=9cm,angle=0,file=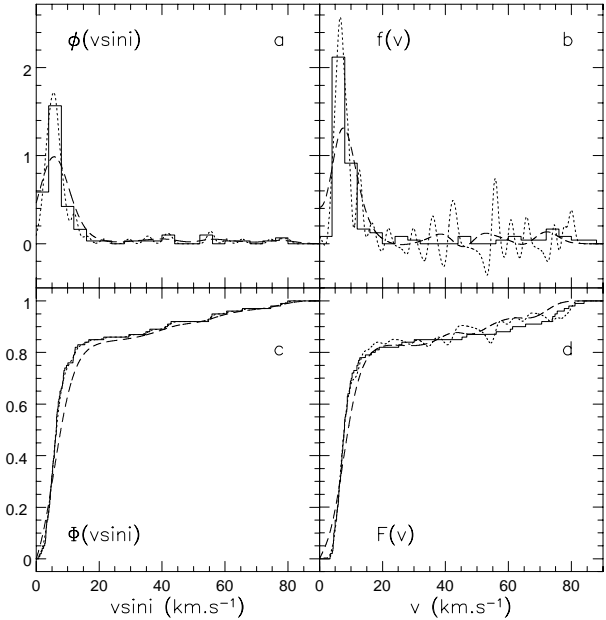}
  \caption{Example of the influence of the parameter $l$ on the inversion 
procedure and the measurement of the rotational velocity distribution for a 
sample of 100 measurements.
{\bf a} solid line histogram displays  a simulated projected velocity 
distribution computed from the $f_0(v)$ distribution showed in 
panel b. The continue estimation  $\phi(y)$  is displayed with dotted line for 
$l=1${\,km\,s$^{-1}$} and 
dashed line $l=4${\,km\,s$^{-1}$}. {\bf b} solid line histogram indicates 
initial velocity distribution $f_0(v)$. 
Dotted and dashed lines show the final result after the Monte-Carlo simulation 
for repectively 
$l=1${\,km\,s$^{-1}$} and $l=4${\,km\,s$^{-1}$}. {\bf c-d} same than  a and b 
panels but the cumulative distributions are displayed.}\label{sigmatest5}
\end{figure}

As a first order test we have run  this simulation  for  a large number of 
measurements
(1000). In all cases, the ``measured'' equatorial rotational velocity  
distribution  $f(v)$,
was equivalent  to the original $f_0(v)$. It means that
the inversion procedure is robust and does not introduce systematic 
biases in the final result.
We proceed by investigating the influence of $l$ and the sample size ($N$) with 
the same Monte-Carlo method. We used various values of $N$ 
between 20 and 200 and two values of $l$: 1 and 4~{\,km\,s$^{-1}$}.  
In Fig.~\ref{sigmatest5} we display the  result of these simulations for $N=100$.

Regardless of the
sample size, the peak of the synthetic $\phi(y)$ distribution is well
fitted using a small value of $l$, while for larger $l$ the agreement is
poor. For the high-velocity tail, where the number of stars
is small, oscillations tend to develop for low-$N$ and low-$l$ values. The
main issue is thus to determine the correct value of $l$, as a function of
$N$, which yields a good fit to the original distribution.
On the one hand, a small value of $l$ leads to unrealistic fluctuations in
the fit of the $\phi(y)$ distributions, especially for the tail of fast
rotators.  On the other, a large $l$ value leads to a much smoother  curve
which does not reproduce well the sharp low-velocity peak. The high-velocity 
tail is better reproduced with a large $l$ value and the details of the
low-velocity peak with a smaller value of $l$ .

Cumulative distributions are less sensitive to the parameter $l$.
For a given rotation value, the frequency
distribution is sensitive to the number of stars in the bin-width $l$
centered on this value, while cumulative distribution depends only on the
sum of projected or equatorial rotational velocity up to this value.  In
all cases, a small $l$, leads to a better match of the cumulative
distributions.

The inversion of both curves ($\phi(y)_{l=1}$ and $\phi(y)_{l=4}$)
is displayed on Fig.~\ref{sigmatest5}b and d.
Statistical fluctuations are amplified by the numerical inversion leading
to important differences between the two curves. For small $l$, the
frequency distribution can even take unphysical negative values in the
high-velocity tail, leading to a local decrease of the cumulative
distribution.

\begin{figure*}[!htpf]
\begin{tabular}{cc}
\psfig{width=8cm,height=8cm,angle=0,file=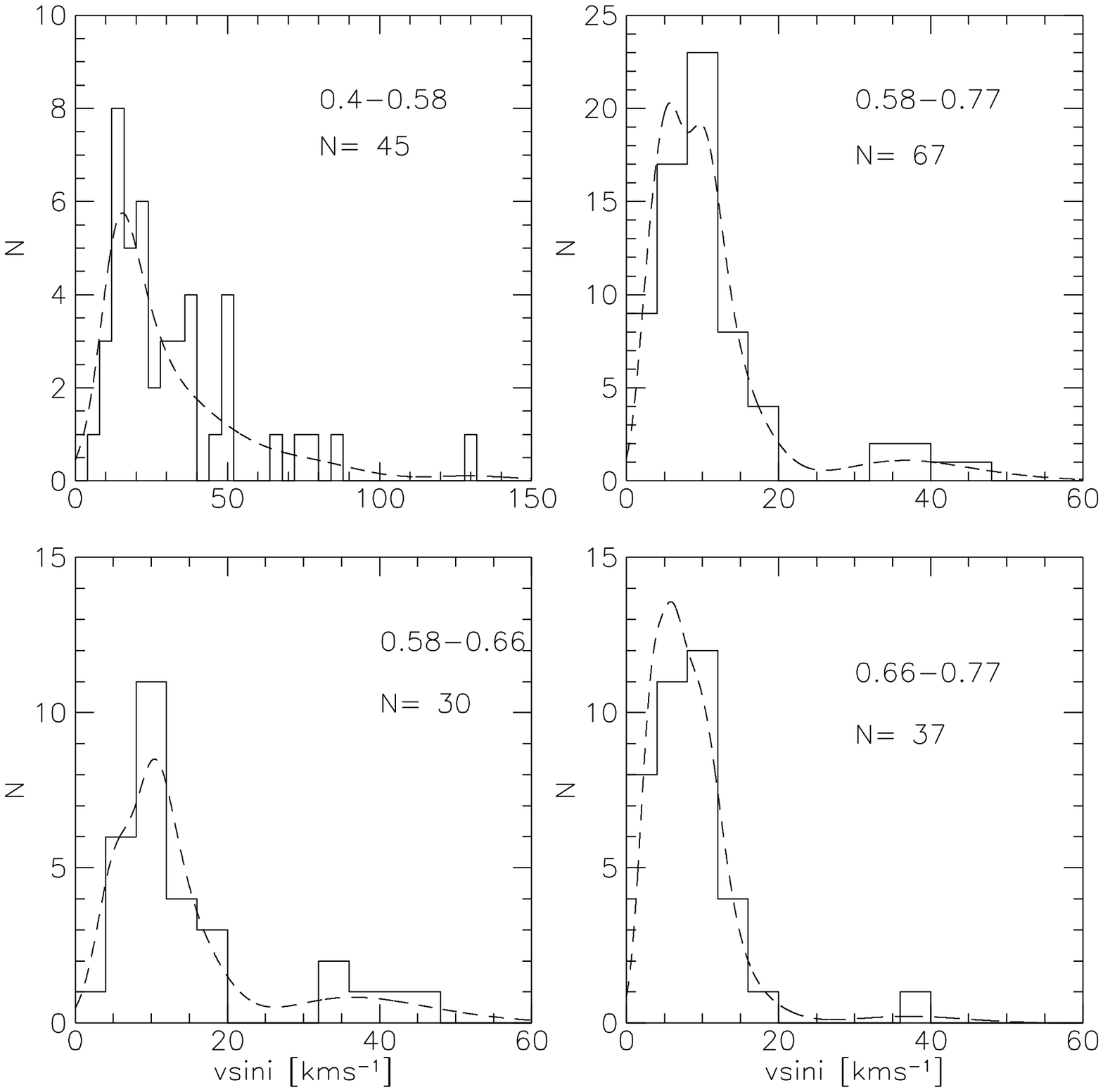} &
\psfig{width=8cm,height=8cm,angle=0,file=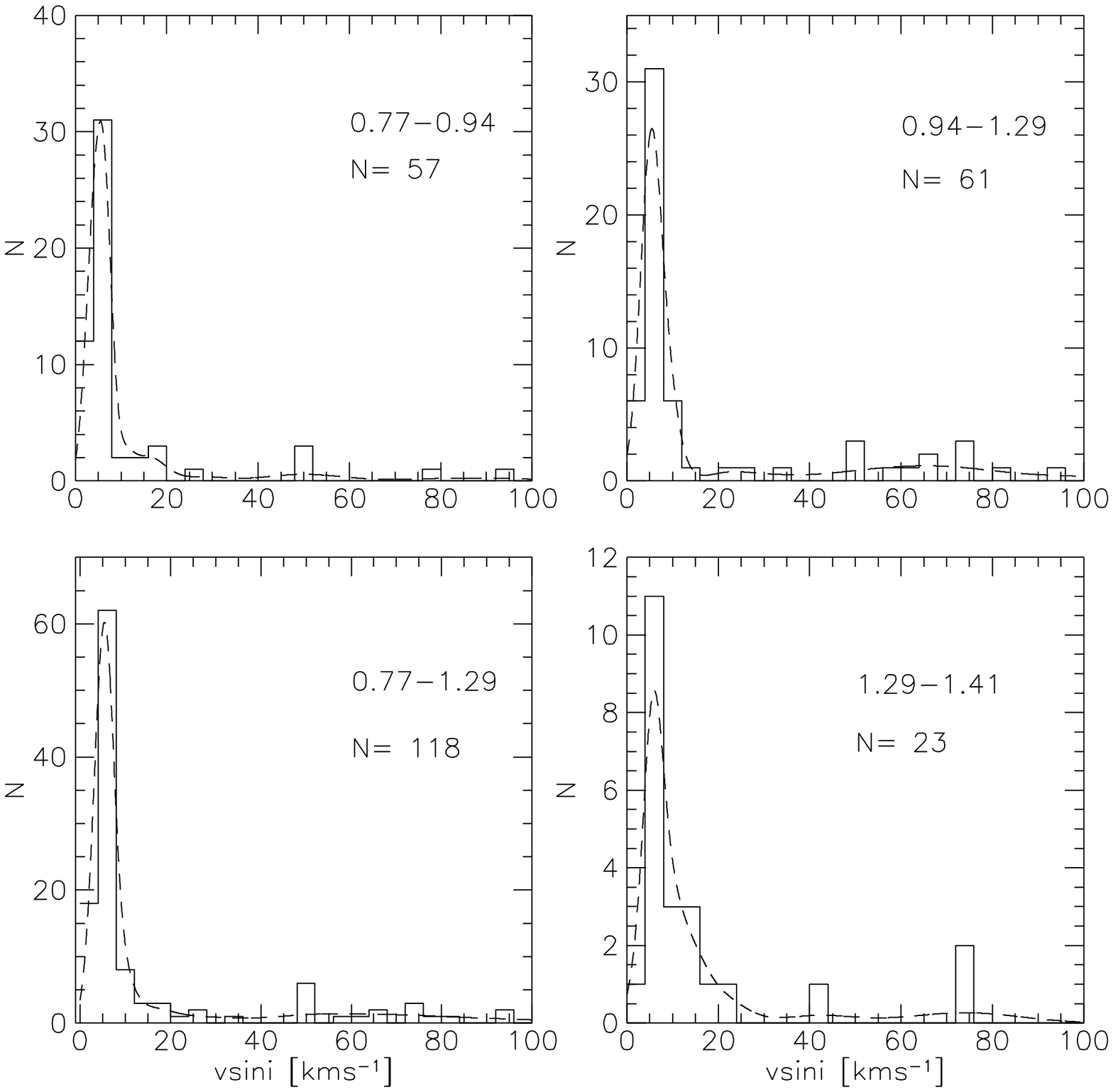} \\
\end{tabular}
\begin{tabular}{cc}
\psfig{width=8cm,height=8cm,angle=0,file=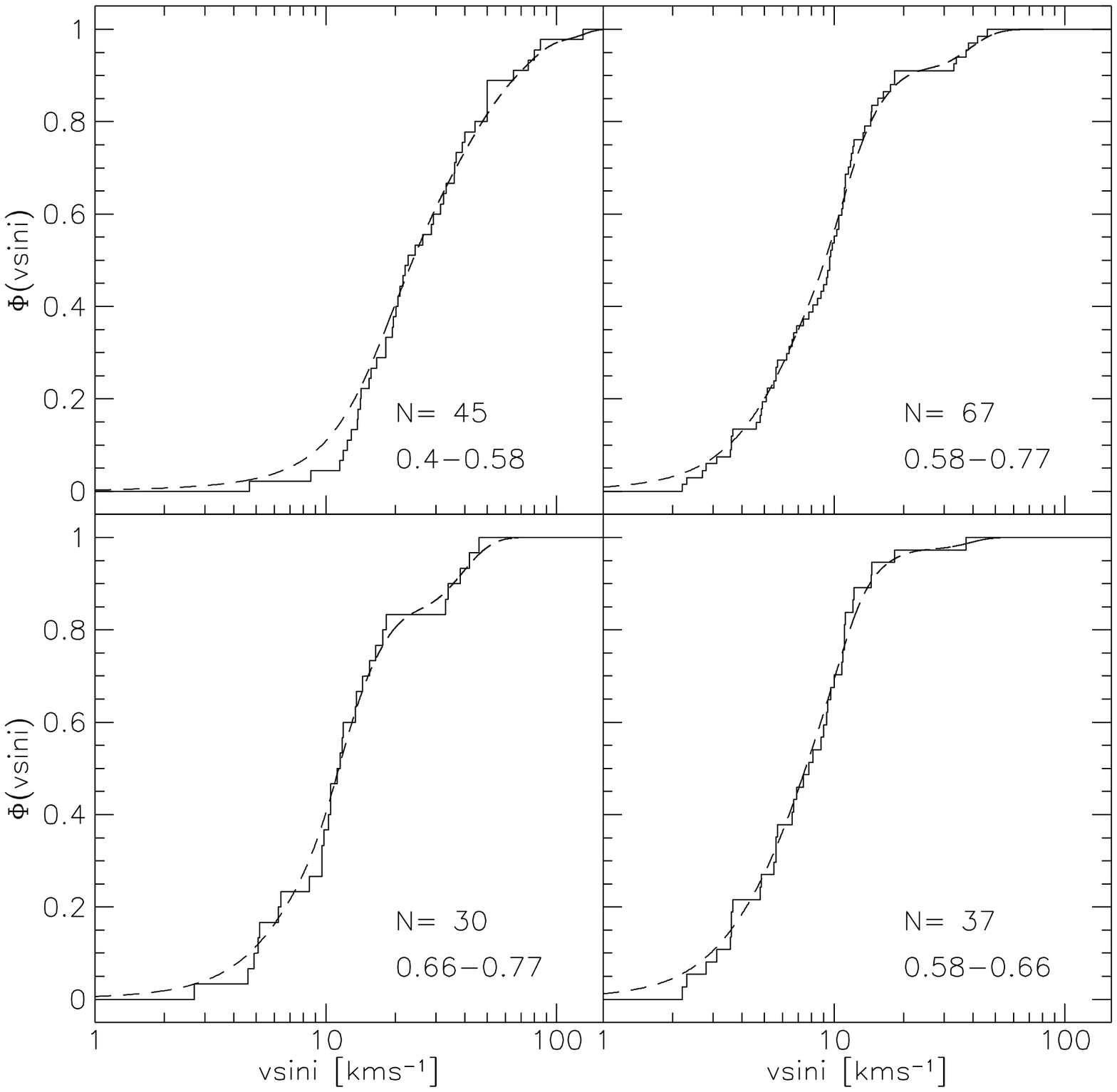} &
\psfig{width=8cm,height=8cm,angle=0,file=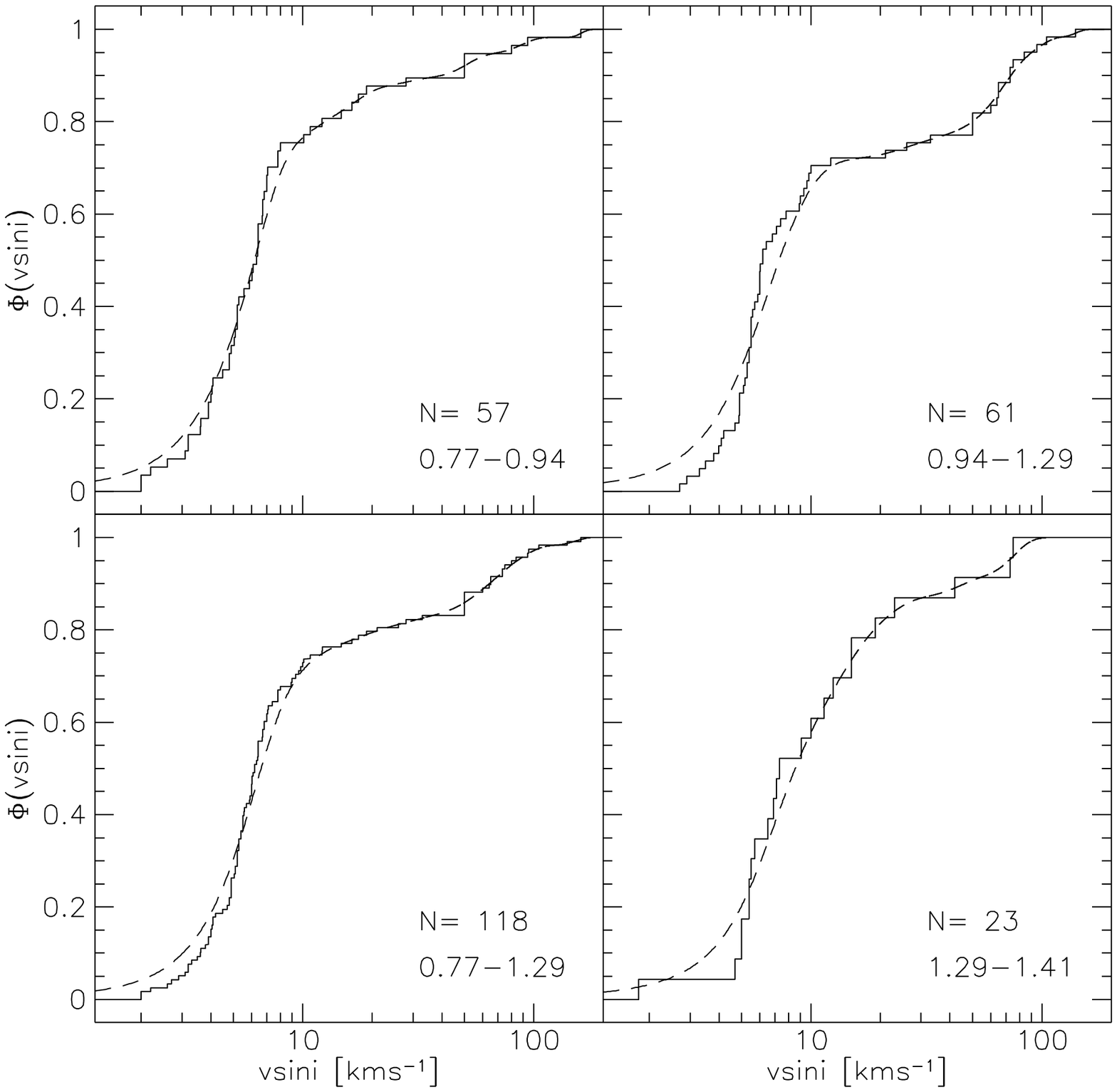} \\
\end{tabular}
\caption{
  {\bf Upper panels} Histograms of observed {$v\sin i$}~ distributions for all
  selected mass domains (displayed in {(B$-$V)$_{0}$}~ ranges) and their fitted 
  continuous distributions (dashed)
  which is used in the inversion process. {\bf Lower panels} similar but
  the cumulative distributions are displayed.}\label{dist1}
\end{figure*}

In summary, two main factors have lead to the  selection of the best value of  
$l$:
the number of measurements and the shape of the {$v\sin i$}~ distribution. In
the case of the Pleiades, the observed distribution shows a sharp peak at low
velocities, and an extending tail to the rapid rotators (see on
Fig.~\ref{dist1}). With  the large numbers of slow rotators and the high
precision of {$v\sin i$}~ measurements one can  use  a small $l$ in this 
domain.  On
the contrary, the small number of fast rotators requires a large value of
$l$. Therefore we decided to use a $l$ parameter which scales upon
{$v\sin i$}~ as  $l\approx 0.2\, v\sin i$,  with a minimum of 
0.8~{\,km\,s$^{-1}$} for
small rotators and a maximum of 10~{\,km\,s$^{-1}$} for large {$v\sin i$}~ 
(these values might slightly vary from one mass bin to another). 
In addition we have imposed a lower limit of 1.5~{\,km\,s$^{-1}$} 
on the value of $l$ corresponding to the
errors on {$v\sin i$}~ measurements for slow rotators.  
Our results  displayed on
Fig.~\ref{dist1} indicates that this choice is fairly good. 

\section{Equatorial rotational velocity distributions in the Pleiades}

The distributions of projected velocities observed in the Pleiades are fitted
and inverted using the inversion procedure described in the previous
section. Since all {$v\sin i$}~ are resolved and we have a complete sample free
of any selection effect up to {(B$-$V)$_{0}$}$=1.3$, the equatorial
velocity distributions for various mass ranges in the 0.6-1.4 {$M_{\odot}$}~ 
domain
can be measured.  Table~\ref{massrange} lists the mass bins within which
the inversion was performed and the corresponding number of measurements.
The sample below 0.6~{$M_{\odot}$}~ ({(B$-$V)$_{0}$}$>1.29$) is not complete 
and is biased
towards the moderate rotators by some {$v\sin i$}~ upper limits.  The 
statistical
results computed for this mass range have to be taken with caution.

\begin{figure*}[!htpf]
  \psfig{width=17cm,height=11cm,file=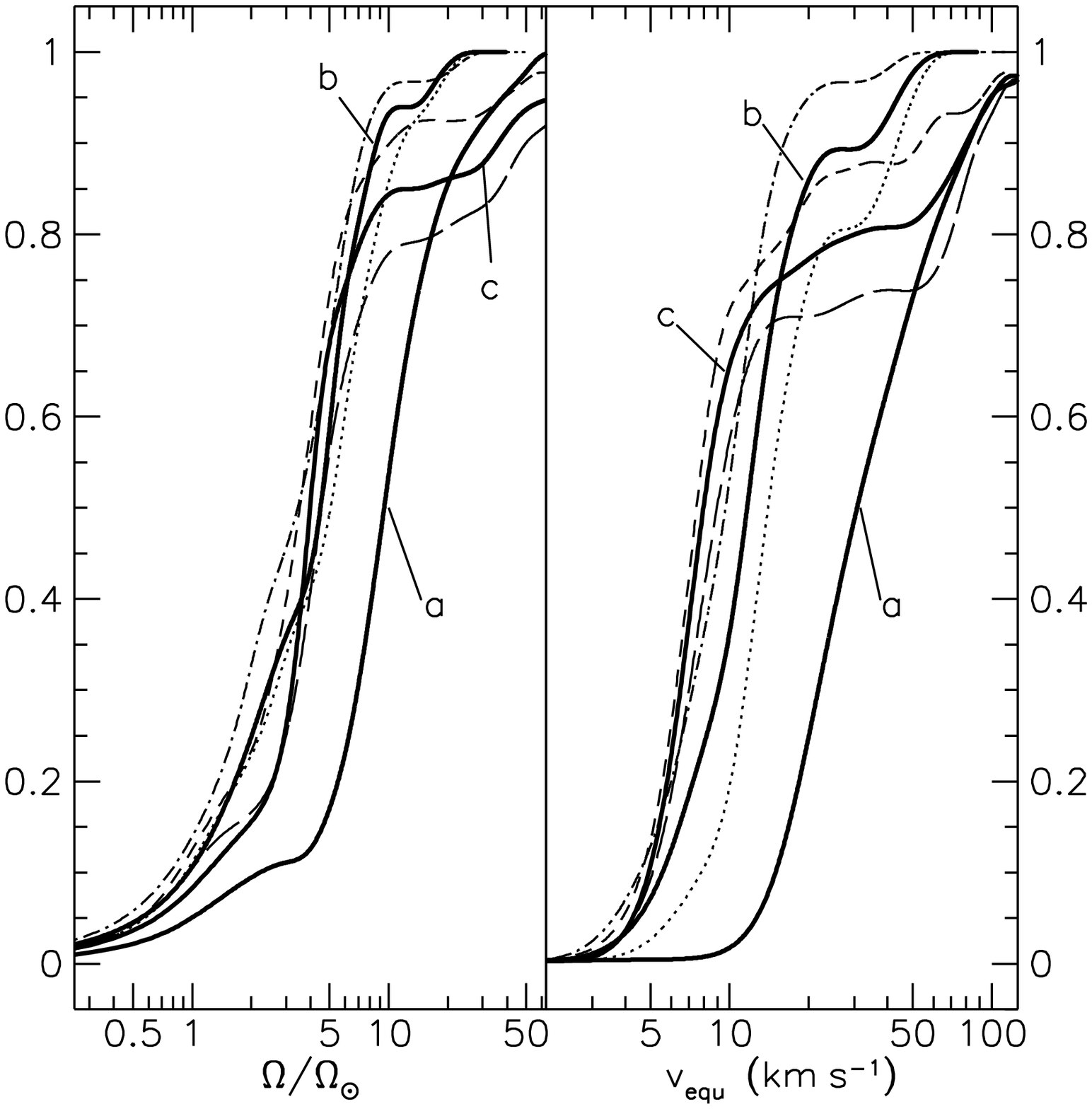}
  \caption{
    Cumulative distributions of the equatorial ({$V_{\hbox{\scriptsize equ}}$}) 
and angular (
    $\Omega/\Omega_\odot$) velocity in Pleiades for various mass
  domains ({$M_{\odot}$}). Wide bins : {\bf
      a} 1.1 -- 1.5, {\bf b} 0.9 -- 1.1, {\bf c} 0.6 -- 0.9. Narrow
    bins:  (dot) 1.0 -- 1.1, (dot--dash) 0.9 -- 1.0, (short dash)
    0.75 -- 0.9, (long dash) 0.6 --0.75.\label{dist2}}
\end{figure*}
 
The cumulative distributions of equatorial velocities for each bin of mass
are displayed in Fig.~\ref{dist2}. They show very different shapes between
each others.  Among the stars with masses above 1.1~{$M_{\odot}$}~ there are no
slow rotators. Below 1{$M_{\odot}$}, the fraction of slow rotators dramatically
increases and most of them have an equatorial velocity less than 
15~{\,km\,s$^{-1}$}.
The smallest fraction of fast rotators ($\geq$ 20 {\,km\,s$^{-1}$}) is measured 
in the
0.9--1.0 {$M_{\odot}$}~ mass bin. But the highest ratio of very slow rotators
($\leq$ 10 {\,km\,s$^{-1}$}) is observed for K dwarfs.  About half of stars 
with a mass
between 0.75{$M_{\odot}$}~ and 0.9{$M_{\odot}$}~ have rotations less than 
7.5{\,km\,s$^{-1}$} or
$\Omega<5\Omega_\odot$.

In order to check if there is any difference between the velocity
distributions in the core and in the corona (stars located at a distance
larger than 80' away from the center), we computed distributions for both
samples separately.  For all the wide mass ranges studied (0.6--0.9, 0.9--1.1, 
1.1--1.5 {$M_{\odot}$})
no significant difference was found between the corona and the core. 
However in the 0.75-0.9 {$M_{\odot}$}~ sub-range  a slight excess of slow 
rotator is
found in the corona compared to the core. We interpret this difference  as 
the result of a contamination of our data by a few non--members.   Indeed,
after having rejected the 2 slowest rotators no significant difference 
remains between the two distributions.
Considering the size of our samples and  the small number of  non-members 
foreseen from the selection process, we expect that the  contamination will not
significantly affect the conclusions that we have derived.

In order to set reliable constraints for the modeling of angular momentum
evolution of stars, we have computed the error bars on the cumulative
distributions due to the finite number of measurements used. For each mass
bin we have first randomly generated a synthetic set of equatorial
rotational velocities equal to the number of {$v\sin i$}~ measurements carried
out and statistically distributed along the inverted distribution $f(v)$.
We have multiplied each rotational velocity by a random value of {$\sin i$}~ in
order to get a simulated set of "observed data". We have finally applied
our inversion procedure to derive a reconstructed $f(v)$ distribution.  By
repeating this process many times (100) we obtain an estimate of the
statistical noise affecting our final distributions. The resulting
1$\sigma$ error bars are listed in Table~\ref{massrange}.

\begin{table*}[!htpf]
\caption{
 Statistics on the distribution of the rotational velocity in Pleiades for
each mass domain selected: $N$ is  the number of stars in the mass bin
considered. In the upper part of the Table, the  percentages of stars  with
velocities less 
than 20, 10, 7.5 and 5 {\,km\,s$^{-1}$} are listed in each mass bin. In the
lower part, the equatorial  velocity $v$ corresponding to various values of the 
cumulative
distribution is indicated. See the text for explanations about the
error bar estimates.}\label{massrange}
\begin{flushleft}
\begin{tabular}{llccccc}
\hline
\noalign{\smallskip}
{(B$-$V)$_{0}$}         & Mass         & N   &\multicolumn{4}{c}{Fraction of 
stars (in \%)  with}\\ 
\noalign{\smallskip}
\cline{4-7}
\noalign{\smallskip}
range        & {$M_{\odot}$}        &     &$v<$20{\,km\,s$^{-1}$}    
&$v<$10{\,km\,s$^{-1}$}    &$v<$7.5{\,km\,s$^{-1}$}  & $v<$5{\,km\,s$^{-1}$}  \\
\noalign{\smallskip}
\hline                           
\noalign{\smallskip}
{\bf 0.4 -- 0.58}  & {\bf 1.1 -- 1.5} & 45 & 24.6$\pm$5.4 & 1.7$\pm$2     & 
0.6$\pm$1.3  & 0.4$\pm$0.7\\
\noalign{\smallskip}
{\bf 0.58 -- 0.77} & {\bf 0.9 -- 1.1} & 67 & 85.8$\pm$4.6 & 36.1$\pm$6.1  & 
21.3$\pm$7.1  & 7.1$\pm$ 3.6\\
0.58 -- 0.66       & 1.0 -- 1.1       & 30 & 74.6$\pm$8.4 & 19.4$\pm$8.6  & 
9.1$\pm$5.7   & 2.7$\pm$ 3.1\\
0.66 -- 0.77       & 0.9 -- 1.0       & 37 & 94.9$\pm$3.7 & 52.9$\pm$10.0 & 
32.3$\pm$9.2  & 12.8$\pm$5.9\\
\noalign{\smallskip}
{\bf 0.77 -- 1.29} & {\bf 0.6 -- 0.9} &118 & 77.3$\pm$3.7 & 65.5$\pm$4.1  & 
44.0$\pm$3.9  & 10.6$\pm$2.9 \\
0.77 -- 0.94       & 0.75 -- 0.9      & 57 & 84.2$\pm$4.1 & 71.6$\pm$5.3  & 
50.2$\pm$5.7  & 13.4$\pm$4.7\\
0.94 -- 1.29       & 0.6 -- 0.75      & 61 & 71.0$\pm$6.2 & 57.4$\pm$5.1  & 
35.3$\pm$4.9  &  9.1$\pm$3.4\\
\noalign{\smallskip}
{\bf 1.29 -- 1.41} & {\bf 0.5 -- 0.6}($^\dagger$) & 23 & 76.4$\pm$9 & 
45.7$\pm$9.5 & 26.5$\pm$7.7 & 5.6$\pm$ 4.9 \\
\noalign{\smallskip}
 \cline{4-7}
\noalign{\smallskip}
 & &     &\multicolumn{4}{c}{velocity $v$ ({\,km\,s$^{-1}$}) such that}\\
\noalign{\smallskip}
 \cline{4-7}
\noalign{\smallskip}
& &        & $F(v)<$10\%  & $F(v)<$20\%  & $F(v)<$50\%  & $F(v)<$90\% \\
\noalign{\smallskip}
 \cline{4-7}
\noalign{\smallskip}
{\bf 0.4 -- 0.58}  & {\bf 1.1 -- 1.5}  & 45 & 14.9$\pm$2.1 & 18.4$\pm$1.9  & 
30.6$\pm$2.7 & 82 $\pm$ 11\\
\noalign{\smallskip}
{\bf 0.58 -- 0.77} & {\bf 0.9 -- 1.1}  & 67 & 5.5 $\pm$1.0 &  7.2$\pm$0.9  & 
11.6$\pm$0.9 & 34 $\pm$ 8\\
0.58 -- 0.66       & 1.0 -- 1.1        & 30 & 7.8 $\pm$1.3 & 10.1$\pm$1.2  & 
14.0$\pm$1.7 & 44 $\pm$ 10\\
0.66 -- 0.77       & 0.9 -- 1.0        & 37 & 4.5 $\pm$1.0 &  6.0$\pm$0.9  &  
9.6$\pm$1.0 & 16.1$\pm$3.3\\
\noalign{\smallskip}
{\bf 0.77 -- 1.29} & {\bf 0.6 -- 0.9 } &118 & 4.9 $\pm$0.4 &  5.8$\pm$0.3  &  
7.9$\pm$0.4 & 80 $\pm$ 13\\
0.77 -- 0.94       & 0.75 -- 0.9       & 57 & 4.6 $\pm$0.5 &  5.5$\pm$0.4  &  
7.5$\pm$0.4 & 53 $\pm$ 23\\
0.94 -- 1.29       & 0.6 -- 0.75       & 61 & 5.0 $\pm$0.5 &  6.1$\pm$0.5  &  
8.9$\pm$3.3 & 89 $\pm$ 13\\
{\bf 1.29 -- 1.41} & {\bf 0.5 -- 0.6}($^\dagger$) & 23 & 5.6$\pm$1.0 & 
6.8$\pm$1.0 & 10.8$\pm$2.0 & 72 $\pm$ 25\\
\noalign{\smallskip}
\hline

\end{tabular}

\end{flushleft}
($^\dagger$) The selected sample in this mass range suffers from incompletness. 
The number of 
slow rotators  is underestimated. The fraction of stars  with 
$v<7.5${\,km\,s$^{-1}$} and $v<5${\,km\,s$^{-1}$} are only lower limit 
estimates.
\end{table*}

\section{Discussion}
\subsection{Comparison with others clusters}

The Hyades cluster is an excellent comparison cluster. It is older, its
members are well known and rotational periods have been measured for
numerous stars over a mass domain starting from late F up to late K (Radick
et al. 1987). From G to early K the rotators show very small scatter per
mass domain, so that it is possible to define a single mass-rotation
sequence and use it as a reference.  Few others clusters have such a
complete sample of rotational data carried out without any selection or
observational bias. In M\,34, IC\,2391 and IC\,2602 recent measurements
have been collected respectively by Jones et al. (1997) and Stauffer et al.
(1997b). The G dwarf samples in these clusters  only include part of
the cluster stars but according to these authors they are free of any
significant observational bias towards fast rotators, such as X-ray 
selection. The contamination of M34 candidate members by field stars 
is likely higher than in IC\,2391 and IC\,2602 and its slow rotators
barely resolved. Nevertheless, because the age of M\,34
is so critical to characterize the evolution of rotation in late-type
dwarfs, we have decided to retain this set of data.  For our analysis we
consider that all these clusters can be used as an age sequence to track
the evolution of the rotation of the slow rotators before and after the
Pleiades age. We use the median of the rotational velocity distribution
($F(v)=50$\%), a kind of upper limit for the slow rotators distribution, as
a statistical indicator of the behavior of this population.  This
indicator, contrary to the mean, does not depend on the exact value of
rotation for the smaller rotators. It is also little affected by the
incompleteness that could arise from a few very-fast rotators which could
have been missed or which have spectral lines too much broadened to measure
reliable {$v\sin i$}. For M34, IC\,2391 and IC\,2602 clusters, the median of the
velocity distributions is estimated from the median of {$v\sin i$}~
distributions divided by $4/\pi$.  Monte--Carlo simulations show that this
rough correction is accurate enough (10\% difference) if the velocity
distribution has a similar shape to that measured in the Pleiades for the
$0.58<${(B$-$V)$_{0}$}$<0.77$  domain.

\begin{figure}[!htpf]
  \psfig{width=8cm,height=8cm,file=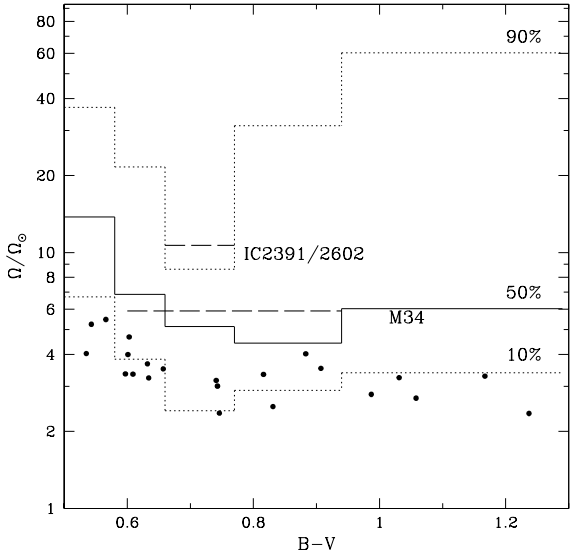}
  \caption{
    Rotational velocity distribution of the Pleiades compared to that of
    the Hyades, M34 and IC2391/2602. The dots represents the individual
    rotation period measurements of Hyades dwarfs converted to angular
    velocity. The median of the rotational velocity distribution for the
    Pleiades $F(<50\%)$ and two border value $F(<10\%)$ and $F(<90\%)$ are
    displayed by solid and dotted lines. Dashed lines indicates the median
    of the velocity distribution for G-dwarfs for both young clusters
    IC\,2391, IC\,2602 and M34 computed from their {$v\sin i$}~ cumulative
    distributions and corrected by $4/\pi$.\label{compclu}}
\end{figure}

The comparison between these clusters is shown in
Fig.~\ref{compclu}. It indicates that
50\% of the Pleiades G and K dwarfs have a rotation rate at most about
twice as large as that of Hyades dwarfs. With an age of 100\,Myr for the
Pleiades (see Basri et al. 1996 for a very complete discussion of this
value), and an age of 600\,Myr for the Hyades, this would lead to
$\Omega(t)\approx t^{-\alpha}$, with $\alpha<0.3$ instead of the canonical
0.5 exponent of Skumanich's (1972) relationship.  Since the G dwarfs and
early K dwarfs do not contract any more beyond 100\,Myr, their angular
momentum evolution is only driven by losses from stellar wind and,
possibly, the resurfacing of the (faster) core rotation. The small value of
the exponent that we find for the Skumanich-type power-law above is
therefore an indication that either the braking of moderate rotators is
less efficient than predicted by the asymptotical Skumanich's relationship
($dJ/dt \propto \Omega^3$) or that some recoupling between the core and
envelope occurs between the age of the Pleiades and the Hyades age, with
the fast spinning core transferring angular momentum to the envelope. 

The 10\% slowest rotators of the Pleiades have rotation rates quite similar
to those of Hyades dwarfs. If we scale the rotational behaviour of moderate
rotators from Pleiades to Hyades to this population of very slow rotators,
we would expect to find some stars in the Hyades with rotation periods similar
to the Sun. Even though this seems to conflict with current data on the
rotation of late-type dwarfs in the Hyades, the relatively small sample
studied by Radick et al. (1987) does not preclude the existence of 10\% or
so extremely slow, solar-type rotators in the Hyades. 

IC\,2391 and IC\,2602 are both very young clusters of about 35\,Myr and
their G dwarfs have just arrived  on the main sequence. The G dwarf sample in
these clusters therefore provides an estimate of the rotational
distribution on the ZAMS. The median rotation rate observed in these
clusters agrees well with the value predicted by Skumanich's relationship
applied backwards in time starting from the Pleiades and assuming solid
body rotation. This suggests that these rotators experience non-saturated
braking\footnote{We refer to  ``saturated'' and 
``non-saturated'' breakings  as the two different 
breaking laws usually used in the angular momentum models for respectively  
fast and slow rotators. 
Discussions and references  regarding the link between the observed saturation 
of the  activity indicators 
for fast rotators (see sect. 5.2) and the  suggested saturation of  the 
breaking laws can be 
found in Bouvier et al. (1997)}, i.e., $dJ/dt \propto \Omega^3$. 
Since the median velocity of
Pleiades and Hyades dwarfs is even lower than that of IC cluster dwarfs,
they too must lie in a regime where the angular momentum losses are not yet
saturated, which occurs at about 25{\,km\,s$^{-1}$}~ for solar type stars 
(see sect.~5.2). 
If a "partially-saturated" braking regime had occurred between the
IC clusters and the Pleiades, we would have observed more fast rotators in
the Pleiades. Similarily, recoupling of the fast rotating core with the
slow rotating envelope earlier than Pleiades is also unlikely since in this
case a faster median rotation would have been measured in the Pleiades.
Altogether, the evolution of the median velocity from the Pleiades to the
Hyades is then best understood as the result of angular momentum from the
core resurfacing into the envelope within this few 100 Myr time span.

The median rotation rate of G and early K dwarfs in the M34 cluster is
slightly higher than the Pleiades one.  According to Jones et al.  (1997),
contamination by field stars can be expected in the M34 sample, which would
underestimates the true median velocity. Since M34 is older than Pleiades
--between 180\,Myr (Meynet et al.  1993) and 250\,Myr (Jones et al. 1997)--,
this result does not fit immediately into the paradigm of the continuous
braking of stellar rotation with time on the main sequence.  The different
ages assigned to M34 by Jones et al. (1997) and Meynet et al. (1993) mostly
result from the use of different reddening corrections: 0.07 in the former,
and 0.11 in the latter. A reanalysis of Geneva photometric measurements and
UBV data tends to favor a higher reddening and the Meynet et al.  (1993)
age of 180\,Myr. The age difference between Pleiades and M34 could thus be
sufficiently small so that the rotational distributions would look somewhat
similar. In addition, there is some uncertainty on the  value of
G dwarf median rotation velocity of M34  due to the small number of 
observations carried out (18). However, the nearly equal median 
velocities measured for
G dwarfs in the Pleiades and in M34 can be readily understood if
resurfacing of core angular momentum into the convective envelope occurs on
a time scale of about 100\,Myr. 

These results thus provide a coherent global picture for the observed
evolution of the median velocity of G dwarfs from 35 Myr to 600 Myr, which
strongly suggests that, for slow and medium rotators, the initially 
fast rotating radiative core transfers
angular momentum to the convective envelope on a timescale of 100--200\,Myr.
This work does not set any constraint on the coupling time for fast 
rotators.   

An alternative to the core-decoupling interpretation would be to
assume that the rotational distribution on the ZAMS differs from cluster to
cluster. The spread in initial angular momentum builds up during the
pre-main sequence as the star is still locked to its circumstellar disk.
Any difference in the disk-lifetime distribution between clusters would
thus lead to different ZAMS velocity distributions. In such a case, the
early velocity distribution would be primarily driven by the physical
parameters of the proto--stellar local environment and not by the age of
clusters.  Moreover the subtle effects of metallicity and binary content
upon angular momentum losses are not yet elucidated. However, there is
currently no evidence that these effects can strongly impact upon the
angular momentum evolution of young low-mass stars. For instance, both
single and binary G and K dwarfs in the Pleiades appear to follow the same
rotational velocity distribution (Bouvier et al. 1997a). Pending evidence
that other processes than those currently included in the models can affect
the angular momentum evolution of young stars, our results favor the
hypothesis of core-envelope recoupling during the early main sequence
evolution.

\subsection{The activity--rotation connection}

Since the pioneering work of Kraft (1967), many studies have been carried
out about the relation between the rotation and the activity of stars.  It
is now obvious that both the chromospheric and the coronal emission is
tightly linked to the stellar rotation and to the depth of the stellar
convection zone --or convective turnover time $\tau_c$-- and such a
relationship is consistent with the qualitative predictions of the
$\alpha\omega$-dynamo magnetic field generation.

The recent ROSAT observations of nearby open clusters have boosted our
knowledge about the X-ray emission of young stars (see Caillault 1996, for
references).  The best way to display the connection between the magnetic
field generation and the stellar activity is probably the Rossby diagram
(activity indicator versus Rossby number $R_0$). Noyes et al. (1984) has
clearly shown, using the Ca\,H-K lines as an activity indicator, that such
a diagram allows one to combine stars with  a wide range of mass without 
significantly increasing the dispersion.

When the Rossby diagram is displayed for dwarf stars with spectral types
between F5 to M5, using ${\log(L_{\hbox{\scriptsize X}}/L_{\hbox{\scriptsize 
bol}})}$ as the coronal activity indicator,
 it shows an obvious discontinuity near $\log(N_R)\approx-.75$. On
the one hand, for the rapid rotators, a "saturation plateau" suggests an
upper limit of about ${\log(L_{\hbox{\scriptsize X}}/L_{\hbox{\scriptsize 
bol}})}\approx-3$. On the other hand, for slow rotators
--Hyades and most field stars--, ${\log(L_{\hbox{\scriptsize 
X}}/L_{\hbox{\scriptsize bol}})}$ steadily decreases when the
Rossby number increases (see for examples Patten \& Simon 1996; Randich et
al.  1996). The exact location of the transition between the saturated and
unsaturated regimes is still ill-defined due to the lack of precise
rotation measurements for rotators in the 5 to 15{\,km\,s$^{-1}$} range.  Most 
of the G
and K dwarfs in the Pleiades have rotation rates less than 20{\,km\,s$^{-1}$} 
and
therefore lie around the transition zone between the "saturation plateau"
and the slow rotators in the Rossby diagram. They thus provide an 
opportunity to fill a scarce populated region of the Rossby diagram.

To build a Rossby diagram with our  Pleiades data, we have converted our
{$v\sin i$}~ measurements into a {$v\sin i$}-based Rossby number: $\log 
(R_0/\sin
i) = \log(P/\tau_c) - \log(\sin i) = \log(P/\sin i) + \log(1/\tau_c)$, where
$P/\sin i$ is derived from {$v\sin i$}~ using the radius--{(B$-$V)$_{0}$}~ 
calibration of
Schmidt-Kaler (1982) and the turnover time $\tau_c$ is computed from
Eq.~4 of Noyes et al. (1984).  The cumulative distribution of
$\log(\sin i)$ is a very steep function around zero with a mean of $-0.13$
and a 90\% probability to have $\log(\sin i)>-.38$.  The spread in the
Rossby diagram induced by the {$\sin i$}~ distribution alone is less than the
spread observed for Hyades and field stars at a given 
${\log(L_{\hbox{\scriptsize X}}/L_{\hbox{\scriptsize bol}})}$ value in this
diagram.  The {$\sin i$}~ statistics therefore merely introduces a bias in the
diagram by systematically shifting the data towards higher Rossby numbers.
The distribution of $\log(\sin i)$ is strongly skewed. Quite often in the
literature, individual {$v\sin i$}~ measurements are transformed into 
{$V_{\hbox{\scriptsize equ}}$}~ by
multiplying {$v\sin i$}~ by the average factor $4/\pi$.  However, this
correction is valid only for the mean rotational velocity and cannot be
applied to individual measurements.

On Fig.~\ref{rossby} the Rossby diagram is displayed for all Pleiades
dwarfs with $0.5<${(B$-$V)$_{0}$}$<1.4$ and X-ray measurements from ROSAT PSPC 
imager.
All known binaries have been rejected. As expected, the Pleiades data lie
around the transition zone. Taking into account the statistical shift and
the spread arising from the {$v\sin i$}--based Rossby number, the Pleiades data
perfectly link the slowly rotating field and Hyades dwarfs and the fast
rotators on the X-ray plateau.  The slope in the non-saturated regime is
about $-2$ in agreement with earlier results and theorical arguments from
Maggio et al. (1987) and Schmitt et al. (1985). In a recent work on
Alpha\,Per, Randich et al. (1996) found a much flatter slope of about $-1$.
This value was derived from a least-square fit based on a limited sample of
slow rotators whose rotation periods were partly estimated from {$v\sin i$}~
measurements. The $\chi^2$-fitting only works on normal random distributed 
data but not on {$\sin i$ distributed data. The net result of this misuse
is to give too much weight  to the few outlying  low $\sin i$ stars, thus 
flattening the slope. Therefore 
the Randich et al. (1996) value for Alpha\,Per is much more a lower limit than 
a real estimate of the slope.

\begin{figure}[!htpf]
  \psfig{width=8.cm,height=8cm,file=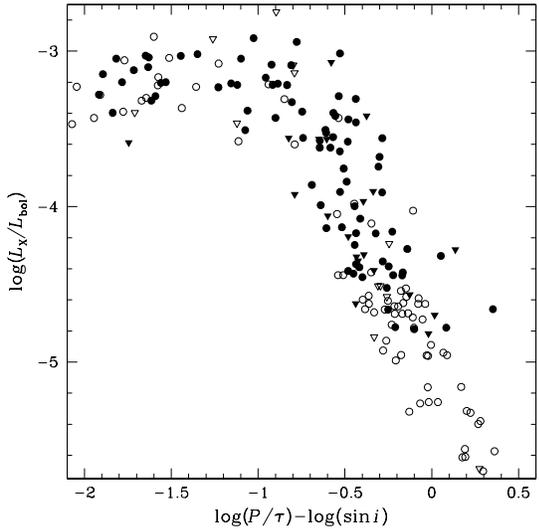}
  \caption{
    Rossby diagram based on {$v\sin i$}~ measurement of Pleiades stars with
    $0.5<${(B$-$V)$_{0}$}$<1.4$ (filled symbols). All the known binaries have 
    been rejected. The X-ray data are from Stauffer et al. (1994); Micela et
    al.  (1996).  Empty symbols represents stars from Alpha\,Per, Hyades
    and field stars whose rotational period has been directly measured
    ($\sin i=1$ in this plot) and X-ray measurements come from ROSAT PSPC 
    observations (data from
    Hempelmann et al. 1995; Randich et al. 1996 and Allain et al. 1997).
    \label{rossby}}
\end{figure}

The transition zone lies at about $\log(P/\tau)=-0.8\pm0.1$, in agreement
with a recent work from Stauffer et al. (1997b). The shape of the Rossby
diagram is the same if we restrict the sample to limited mass ranges.
The only effect is to shift the whole set of data towards the saturation
plateau for stars with smaller masses. This effect is easily understood as
these stars have deeper convective zone --and then larger $\tau_c$-- and
include more fast rotators than higher mass stars. In conclusion, the
Rossby diagram provides a coherent description of the rotation--activity
connection in the mass range 0.5--1.2{$M_{\odot}$}. It shows that the saturation
phenomenon is tightly linked to the stellar mass. Such a mass-dependent
saturation effect had been previously suggested by Collier Cameron \& Li
(1994) and Barnes \& Sofia (1996) and is confirmed here by Pleiades data.

\section{Conclusions}
  
From a complete and unbiased set of {$v\sin i$}~ measurements, we have computed
the distribution of equatorial velocities in the Pleiades for various
mass ranges between 0.5 and 1.5{$M_{\odot}$}.  Comparison with the
distribution of rotational velocities in the Hyades, M\,34, IC\,2391 and
IC\,2602 yields a coherent picture for the angular momentum evolution of
the convective envelope.

The comparison with the younger clusters IC\,2391 and IC\,2602 suggests
that most Pleiades G--dwarfs are in an unsaturated braking law regime.  The
rotational evolution of moderate rotators in early stages is in agreement
with a solid body rotation driven by Skumanich's relationship. The
relationship between {$v\sin i$}~ and X-ray emission indicates that the
transition between saturated X-ray emission and its steady decrease with
rotation occurs at about $P=2$\,d or {$v\sin i$}$=25${\,km\,s$^{-1}$} for 
solar-type stars.
About 10\% of Pleiades G dwarfs lie in the saturation domain. 

The comparison with older cluster such as M\,34 and the Hyades strongly
suggests that angular momentum of slow and moderate rotators is transported 
from the fast rotating core
to the convective enveloppe on a time scale of about 100--200\,Myr on the
early main sequence.  An alternative intepretation would call for intrinsic
differences in the distribution of initial angular momenta from clusters to
clusters, which is not currently supported by observations. 

The advent, in the next decade, of multi-fiber spectrographs on large
telescopes will offer the possibility of determining the {$v\sin i$}~
distributions of more remote clusters of various ages and abundances as
well as the rotational properties of very low-mass stars 
(0.1--0.5{$M_{\odot}$}).
This will undoubtedly improve our understanding of the rotational evolution
of young stars and provide new clues to the physical mechanisms responsible
for angular momentum transport in stellar interiors.

\begin{acknowledgements}
We are gratefull to S. Udry who  kindly provided us his help for all operations 
with the {{\footnotesize CORA\-VEL}}~ data\-base and to J. Stauffer, our 
referee, for his  comments  and his useful suggestions to improve the quality 
of the english. D.Q. acknowledges support from the swiss FNRS.
\end{acknowledgements}


\newpage

\end{document}